\documentclass[11pt]{article}
\usepackage{amsmath}
\usepackage[dvips]{graphicx}
\usepackage{amssymb}
\usepackage{cite}
\newcommand{\be}{\begin{equation}}
\newcommand{\ee}{\end{equation}}
\newcommand{\bear}{\begin{eqnarray}}
\newcommand{\eear}{\end{eqnarray}}
\newcommand{\tanb}{\tan \beta}

\newcommand{\vev}[1]{\left\langle #1\right\rangle}

\newcommand{\goes}{\rightarrow} 
 
\newcommand{\GeV}{\; \mathrm{GeV}} 
\newcommand{\TeV}{\; \mathrm{TeV}} 
\newcommand{\lapproxeq}{\lower .7ex\hbox{$\;\stackrel{\textstyle  
<}{\sim}\;$}} 
\newcommand{\gapproxeq}{\lower .7ex\hbox{$\;\stackrel{\textstyle  
>}{\sim}\;$}} 
\newcommand{\stackdown}[2]{\lower 1.4ex\hbox{$\;\stackrel{\textstyle{#1}}  
{\scriptstyle{#2}}\;$}}
\newcommand{\beq}{\begin{equation}} 
\newcommand{\eeq}{\end{equation}} 
\newcommand{\bea}{\begin{eqnarray}}
\newcommand{\eea}{\end{eqnarray}}

\newcommand{\ori}{\hspace{.1in}}
\newcommand{\fp}{f^{\prime}}

\makeatletter 
\def\slash{\@ifnextchar[{\fmsl@sh}{\fmsl@sh[0mu]}} 
\def\fmsl@sh[#1]#2{% 
    {\@fmsl@sh\displaystyle{#1}{#2}}% 
    {\@fmsl@sh\textstyle{#1}{#2}}% 
    {\@fmsl@sh\scriptstyle{#1}{#2}}% 
    {\@fmsl@sh\scriptscriptstyle{#1}{#2}}} 
\def\@fmsl@sh#1#2#3{\m@th\ooalign{$\hfil#1\mkern#2/\hfil$\crcr$#1#3$}} 
\makeatother 
\begin{document} 
\begin{titlepage} 
%%%%%%%%%%% 
\begin{flushright} 
%\parbox{7.4cm}
%%  arXiv:yymm.nnnn [hep-ph] \\           
\end{flushright} 
%%%%%%%%%% 
\vspace*{5mm} 
\begin{center} 
{\large{\textbf {Refining the predictions of supersymmetric $CP$-violating 
models: \\ A top-down approach 
}}}\\
\vspace{11mm} 
{\bf M.~Argyrou} $^{1}$, {\bf A.~B.~Lahanas} $^{1}$,
and 
{\bf V.~C.~Spanos} $^{2}$   \\
\vspace*{6mm} 
 $^{1}$ {\it University of Athens, Physics Department,  
Nuclear and Particle Physics Section,   
GR--15771  Athens, Greece\\}
%%%%%%%%%%%%%%%%%%%%%%%%%%%%%%%%%
\vspace*{4mm} 
 $^{2}${\it  University of Patras, Department of Physics, GR-26500 Patras, Greece}
\end{center} 
\vspace*{2cm} 
%%%%%%%%%%%%%%%%%%%%%%%%% 
\begin{abstract}
We explore in detail the consequences of the CP-violating phases residing in the 
supersymmetric and soft SUSY breaking parameters in the approximation that family 
flavour mixings are ignored. We allow for non-universal boundary conditions and in 
such a consideration the model is described by twelve independent CP-violating 
phases and one angle which misaligns the vacuum expectation values (VEVs) of the 
Higgs scalars. We run two-loop renormalization group equations (RGEs), for 
all parameters involved, including phases, and we properly treat the minimization 
conditions using the one-loop effective potential with CP-violating phases included. 
We show that the two-loop running of phases may induce sizable effects for the electric 
dipole moments (EDMs) that are absent in the one-loop RGE analysis. Also important 
corrections to the EDMs are induced by the Higgs VEVs misalignment angle which are 
sizable in the large $\tan \beta$ region.
Scanning the available parameter space we seek regions compatible with 
accelerator and cosmological data with  emphasis on rapid  neutralino annihilations 
through a Higgs resonance.  
It is  shown that large CP-violating phases, as required in  Baryogenesis scenarios, 
can be tuned to obtain agreement 
with WMAP3 cold dark matter constraints, EDMs and all available accelerator data,  
in extended regions of the parameter space which may be accessible to LHC.
\end{abstract}
\end{titlepage} 
% 
%\newpage 
%\baselineskip=18pt 
\baselineskip=15pt  
%%%%%%%%%%%%%% Paper body %%%%%%%%%%%%%%%%%%%%%%%%%
%%%%%%%%%%%%%%%%%%%%%%%%%%%%%%%%%%%%%%%%%%%%%
\section{Introduction}
Supersymmetry (SUSY) seems to be an indispensable ingredient of Superstring theories 
and supersymmetric extensions of the Standard Model (SM) have attracted the interest 
of physicists for more than two decades or so. Supersymmetric models are the only known 
extensions of the SM that are renormalizable field theories,  
bearing therefore the virtue that radiative corrections can be put 
under control and definite predictions can be made. On the other hand it 
is well known that these models are characterized by many arbitrary parameters, 
even in their most simplified versions, and therefore additional theoretical 
assumptions have to be invoked to lessen their number and build less
 proliferated models. The minimal supersymmetric standard model (MSSM) 
has the minimal physical content and it needs 124 parameters, a large number 
indeed.  These are reduced to  much fewer in the supergravity scenarios (mSUGRA),
 in which universal boundary conditions hold, which can be further reduced if 
additional assumptions are made, as for instance absence of generation mixings 
in the supersymmetric sector at the tree level and/or absence of CP-violating
 phases in the  supersymmetric parameters which in the minimal much studied 
versions of SUSY are switched off. For a review on the availability and the 
description of the various models see Ref. \cite{Chung:2003fi} and references 
therein. A thorough account of the parameters describing the supersymmetric models
 and the issue of the CP-violating phases can be found in \cite{Haber:1997if}.

Supersymmetric models with CP-violating phases, other than this occurring in 
the Cabibbo-Kobayashi-Maskawa matrix (CKM), have been extensively studied in the past 
\cite{EDM1,EDM2,EDM3,EDMrge,Dimopoulos:1995kn,masiero}. 
In the mSUGRA models there are only two observable phases, in addition to the 
CKM phase, which are tightly constrained by the EDM data. In more general cases 
however, when universal boundary conditions are abandoned, the number of phases 
is increased opening new possibilities that greatly affect phenomenology. 
The CP-violating phases residing in the supersymmetric parameters produce
 not only new phenomena, absent in minimal models, but also affect 
CP-conserving quantities, like the mass spectrum for instance, 
or have large impact on various particle processes being therefore 
of relevance in collider searches \cite{Brhlik:1998gu}. 
The reconstruction of the soft supersymmetric Lagrangian from 
experimental data, including cases where CP-phases are present 
is addressed in many works \cite{reconstruct}.

The constraints imposed by the EDM data of neutron \cite{Harris:1999jx},  
$|d_n| < 6.5 \times 10^{-26}\; e \cdot cm$, the EDM of 
electron~\cite{Regan:2002ta} deducted  from measurement of the 
corresponding EDM of thallium, $|d_e| < 1.6 \times 10^{-27}\; e \cdot cm$, 
or diamagnetic atomic systems such as Mercury (Hg-199) \cite{Romalis:2000mg},  
$|d_{Hg}| < 2.1 \times 10^{-28}\; e \cdot cm$, 
 are very tight restricting CP-violating phases to be unnaturaly small. 
This problem  is termed as the supersymmetric CP-problem. 
The phenomenological constraints imposed by measurements of the electric 
dipole moments has been the subject of numerous works \cite{EDM1,EDM2,EDM3,
EDMrge,EDMcancel1,EDMcancel2,EDMcancel2b,CPdirect,
EDMcancel3,Hg199,EDMcancelstring2,EDMcancelstring,EDMsfermion,Pilaftsis:1999td,
Lebedev:2002ne,gammaHiggs,WWloop,other2loop,exactsplit,Giudice:2005rz,
olive3loop,EDM2001,Pospelov:2005pr,Erler:2004cx}~
{\footnote{For a thorough review concerning the role of dimension five and 
six operators affecting the EDMs of atomic systems and their link 
to EDMs of neutron and electron see \cite{Pospelov:2005pr}. 
For a general review concerning low energy tests of the weak 
interactions including EDMs see \cite{Erler:2004cx}.
}}.  To forbid overproduction of EDMs one may assume that the masses of 
superpartners are heavy enough beyond the reach of LHC. In other approaches
 special mechanisms are invoked, such as the cancellation mechanism 
\cite{EDMcancel1,EDMcancel2,EDMcancel2b,EDMcancel3,EDMcancelstring2}, 
in which contributions of the various Feynman graphs involved delicately 
cancel each other to render EDMs of neutron and electron small within their 
experimental limits. However even in this case the limits imposed by the EDM 
of Mercury atom are hard to satisfy \cite{Hg199,EDMcancelstring}.
One may cure the situation by lifting the sfermion masses \cite{EDMsfermion} 
but then two-loop contributions are not suppressed \cite{Pilaftsis:1999td,Lebedev:2002ne}.  
The EDMs have been studied beyond the one-loop order. 
The Barr-Zee type \cite{Barr:1990vd} two-loop supersymmetric contributions 
have been thoroughly studied 
\cite{gammaHiggs,WWloop,other2loop,exactsplit,Giudice:2005rz} and yield 
sizable contributions which do not decouple in the limit of heavy sfermion 
masses. The same holds even for some three-loop contributions arising from 
the running of the renormalization group equations (RGEs) which induce
 sizable phases to the gaugino masses \cite{olive3loop}.

The effect of supersymmetric phases in Higgs sector has been extensively
 studied~\cite{HiggsCP1,HiggsCP2,HiggsCP3,HiggsCP4,HiggsCP5,HiggsCP6,HiggsCP7}  
 and will not be repeated here;
for a review and see \cite{Kraml:2006ga}. We merely state that 
their couplings to other particles differ from the CP-conserving case 
and that the CP-odd Higgs mixes with the CP-even mass eigenstates. 
Therefore, except their involvement to EDMs, in principle, they  also affect neutralino 
relic densities and thus they are  important for cosmological considerations.

Large phases residing in supersymmetric parameters are welcome for 
Baryogenesis which can occur either through Leptogenesis \cite{Fukugita:1986hr} 
or through a strong first order electroweak phase transition
(for reviews see: \cite{EWBreview}). Squark and slepton driven Baryogenesis 
requires a light stop, with mass  $120 \GeV < m_{\tilde t} < m_t$, and in conjunction 
with the fact that the phase transition becomes  too weak for Higgs masses larger 
than $\sim 120 \GeV$ it leaves a narrow window for successful baryon 
asymmetry \cite{lightstop,Carena:1997ki,lighthig,nonpert}. 
Higgsino and Gaugino driven baryogenesis \cite{charEWB,neutrEWB} 
is an alternative. This effect is resonantly enhanced when the 
Higgsino mixing parameter $\mu$  is of the same order with the 
gaugino masses $M_{1,2}$, \cite{resonant}. The relevant CP-phases 
are $arg(\mu M_{1,2})$ and values in the range $\sim 10^{-2}$ are 
adequate to produce the observed baryon asymmetry of the universe if $|\mu| \simeq |M_{1,2}|$.

During the last years the WMAP3 \cite{Bennett:2003bz,Spergel:2003cb,Spergel:2006hy} 
and SDSS \cite{Tegmark:2003ud} precise determination of the cosmological parameters 
has stimulated new interest and, in conjunction with the constraints put by the 
accelerator data, it points to a better understanding and more thorough treatment 
of supersymmetric models that violate CP. The Cold Dark Matter (CDM) relic density 
implied by the latest WMAP3 data \cite{Spergel:2006hy} lies in the range 
$\Omega_{CDM} h_0^2 = 0.1045^{+0.0072}_{-0.0095}$
%%%%%%$\;0.0950< \Omega_{CDM} h_0^2 < 0.1117\;\; (1 \sigma)$ 
imposing severe constraints on  CP-conserving supersymmetric models 
(for a review see \cite{Lahanas:2003bh} ). 
The importance of the CP-phases in conjunction with Dark Matter (DM) observations 
has been the subject of many works\cite{Falk:1995fk}-\cite{Belanger:2006qa},
\cite{neutrEWB}. In these works the effect of the phases on the neutralino 
relic abundance is discussed observing the constraints put by accelerators 
in various supersymmetric scenaria. For a recent review 
see \cite{Belanger:2006qa,Belanger:2006pc}. 

\renewcommand{\labelenumi}{\roman{enumi}.}
\renewcommand{\labelenumii}{\roman{enumi}. \alph{enumii}}

In this work we focus on mSUGRA-type CP-violating models with 
minimal flavour violation (MFV) and seek regions of the parameter 
space which are compatible with cosmological and EDM constraints 
and all available accelerator data. In particular we refine the 
analyses of previous works by taking into account: 
\begin{enumerate}
\addtolength{\itemsep}{-.1\baselineskip}
\item The two-loop renormalization group running of all 
phases included which may induce sizable effects at low energies 
having large impact on EDMs. We show that even small phases at the 
unification scale are responsible of inducing large corrections saturating 
in some cases the experimental limits put on EDMs. Such an effect was 
studied in \cite{olive3loop} for the particular case of the trilinear soft 
scalar coupling, whose phase affects the phases of the gaugino masses.
 However other phases with phenomenological interest, notably the gluino phase, 
may influence EDMs, as we shall show, inducing non-vanishing phases for 
the remaining soft parameters. 
\label{it1}
\item Full treatment of the one-loop minimization conditions 
in the presence of CP-violating sources. The one-loop effective potential 
depends on these phases and correct treatment, taking into account all  
one-loop contributions, shows that a misalignment of the Higgs vacuum 
expectation values (VEVs) is induced which is phenomenologically important. 
It is worth noting that the relative angle between the Higgs VEVs cannot be 
rotated away, even if the Higgs mixing parameter is taken real, by proper 
$U(1)$ Peccei-Quinn or R-symmetries. The appearance of this phase is due 
to one loop corrections of the effective potential. The importance of 
this misalignment angle for phenomenology has been pointed out in previous 
works \cite{Demir:1999hj,Demir:1999zb}. Here we refine previous analyses 
and in conjunction with i) we show that it induces phenomena which although
 in principle small may have large impact especially on the EDMs and the relic 
density of the lightest supesymmetric particle (LSP), which is assumed to be the
lightest  neutralino. 
\item The effect of strong interaction phases which 
in principle do not affect electron's EDM at one-loop,
 such as the gluino phase. We show that, due to two-loop 
RGE running, this phase may induce CP-odd invariant phases 
at low energies, with important consequences for EDMs which 
are absent in the one-loop analysis. Such a phase is phenomenologically 
interesting since it is observable in gluino production and it is known 
to affect  neutralino relic densities indirectly through its influence 
on the bottom mass corrections 
\cite{Gomez:2004ef,Gomez:2004ek,Gomez:2004yv,Argyrou:2004cs,Gomez:2005nr}.  
In this work we also argue that the appearance of such a phase may affect 
the top-down approach of the renormalization group running since it may give
 rise to large corrections for the bottom Yukawa couplings having as a 
consequence the appearance of Landau poles \cite{Nath:1995eq} in the large 
$\tan \beta$ regime.  
\item The updated experimental value of the top mass 
is $m_t=171.4\pm 2.1  \GeV${\footnote{
The central value $m_t=171.4 \GeV$~\cite{Brubaker:2006xn} of the top mass has slided down by 
$0.5  \GeV$ according to more recent $CDF$ and $D0$ analyses~\cite{:2007bxa}.
Throughout this work the value $m_t=171.4 \GeV$ is used.}}, 
which affects the location and shape of the cosmologically allowed 
funnels which open in regions where an LSP pair annihilates through 
a Higgs resonance~\cite{funnels}. It is known that these funnels are
 very sensitive to the input top  (and bottom) quark mass~\cite{like}. 
\end{enumerate}

In our treatment  we follow a top-down approach, which is the appropriate 
handling if the low energy physics has its origin at Planckian energies. 
In this sense CP-violating phases are not given at the Electroweak scale 
but are extracted after a two-loop running of the 
real and imaginary parts of all parameters involved, which are inputs at the unification scale. 
Under these circumstances it is interesting to explore where, and under which 
circumstances, cosmologically allowed regions naturally show up, 
in which the EDM constraints are satisfied, for large phases 
which are relevant for Baryogenesis and other phenomenological issues. 
In our study we pay special attention to the LSP pair annihilation 
through a Higgs resonance which is one of the leading mechanisms 
to produce the right amount for neutralino CDM
especially in the large $\tan \beta$ regime. 

In doing so, we have developed a Fortran numerical algorithm, 
which treats CP-violating sources in the MFV scenario in the 
approximation of neglecting generation mixings from the CKM matrix, 
whose effects are known to be small{\footnote{
It is customary in literature to term as MFV those models that allow 
generation mixings only from the CKM matrix.}}. 
In this scheme soft masses and trilinear soft couplings are assumed diagonal in family space. 
In our study we discuss in detail the issues associated with the mass predictions 
of such models, focusing on a two-loop RGEs running and the effect of the  
CP-phases on quantities that greatly affect the renormalization group flow and 
the extracted low energy parameters. As has been already remarked, in our analysis
 electroweak radiative symmetry breaking is enforced with all one-loop 
contributions to the effective potential duly taken into account when 
CP-violations are switched on. 

This paper is organized as follows. In Section 2 we give a detailed 
account of the role of the phases in conjunction with the class of models studied in this work.  
In Section 3 we discuss all subtleties associated with the phases that are involved 
and discuss all pertinent formulas which have a large impact on  the numerical analysis. 
In Section 4 we discuss the importance of the energy dependence of the phases involved and 
in Section 5 we discuss the constraints arising from EDMs and 
Cosmology and present our main results. We end up with the conclusions in Section 6 and give a summary of our results.  
%%%%%%%%%%%%%%%%%%%%%%%%%%%%%%%%%%%%%%%%%%%%%
\section{Description of the Model}
In the softly broken supersymmetric theory the Lagrangian is split as 
$$
\cal{L}\;=\; {\cal{L}}_{\mathrm{SUSY}} + {\cal{L}}_{\mathrm{soft}} \; \; ,
$$
where all information concerning supersymmetry breaking is encoded in the 
soft part ${\cal{L}}_{\mathrm{soft}}$. 
Proper U(1) and R-transformations of the multiplets involved may be used 
to eliminate some of  phases residing in the parameters describing this  
Lagrangian leaving aside a number which cannot be further rotated away. 
These phases consist an additional set of arbitrary parameters with 
important phenomenological consequences. There are several works 
describing the situation in extensions of the MSSM, in which  the
 presence of such phases is taken into account and their phenomenological 
implications are discussed. 
In this work we study MFV versions of supersymmetric models assuming that 
soft masses and trilinear couplings residing in ${\cal{L}}_{\mathrm{soft}}$ 
are family blind. For brevity this class of models we shall coin CPMSSM.  
In such models the effects of the 1st and 2nd generation Yukawa couplings
 in the running of the renormalization group equations (RGE) is small and 
can be safely neglected from the analysis. 

To start with, it may help to recall the basic field transformations used 
in order to eliminate the redundant phases of the various parameters involved. 
MSSM based models have $U_Q(3)\times U_{U_C}(3)\times U_D(3)$ and $U_L(3)\times U_{E_C}(3)$ 
global symmetries acting on the quark and lepton multiplets which can be
 used to eliminate redundant phases and real parameter from the quark and 
lepton Yukawa couplings. At the end six real quark masses, three CKM angles
 and one CP-violating CMK phase are left which are measurable. Thus, 
in the approximation that the effect of the generation mixing is 
neglected we deal with real Yukawa couplings which are diagonal in 
family space. However in the limit that the Higgs multiplet mixing parameter $\mu=0$ and 
the soft terms are neglected additional symmetries exist. 
These are Peccei-Quinn (PQ) global ${U(1)}_{PQ}$ symmetries 
and R-symmetries $U_{R}(1)$ under which the bosonic and
 fermionic components of the multiplets involved are 
transformed by appropriate phase factors. Under ${U(1)}_{PQ}$ 
the Higgs multiplets $H_{1,2}$ have charge $1$, all quark and 
lepton multiplets have charge $-1/2$ and the vector multiplets 
are neutral, that is carry zero charge. On the other hand the 
R-charge of these multiplets is $1$ for the scalar Higgses, 
$1/2$ for the scalar partners of the quarks and leptons and 
zero for the vector bosons. 
The corresponding Higgsinos, quark and lepton fermions have charges by
one unit less while gauginos carry R-charge +1.
In the literature are used instead the symmetries ${U(1)}_{PQ}$ and ${U(1)}_{R-PQ}$. 
 Under the latter the scalar Higgses carry zero 
charge. A useful way to keep track of the changes implemented by 
these transformations is to assume that the
Higgs mixing parameter $\mu$, the Yukawa couplings $h_{t,b,\tau} ... $, 
the gaugino masses $M_a$, the Higgs scalar mixing parameter $m_3^2$, 
the trilinear couplings $A_{t,b,\tau} ...$  and the squark and 
slepton masses squared $m^2_{\tilde{q},\tilde{l}}$ are spurion 
fields transforming in the way shown in Table 
\ref{tablechar},  so that the Lagrangian $\cal{L}$ is kept invariant. 
Phrased in another way, if a ${U(1)}_{PQ}$ or ${U(1)}_{R-PQ}$ 
transformation is carried out on the fields involved, 
i.e. $f \rightarrow exp(iQ_f \alpha) f$, the parameters 
in the transformed Lagrangian appear multiplied by 
$exp(-i Q \alpha)$, with $Q$ the charge shown in  
Table~\ref{tablechar}. It is apparent that 
${U(1)}_{PQ}$ affects only $\mu, m_3^2$, under which both 
have the same charge, and ${U(1)}_{R-PQ}$ affects 
$\mu$, the gaugino masses and trilinear couplings. 
None of these affects the Yukawa couplings and the soft 
scalar masses which means that if real in one basis they 
remain real after ${U(1)}_{PQ}$ or ${U(1)}_{R-PQ}$ transformations. 
In a particular basis one exploits the aforementioned 
transformations to rotate away redundant phases of the parameters 
involved. Which ones is a matter of convention. However certain 
combinations of phases are invariant under these PQ and
 R-transformations and all physical quantities depend on 
linear combinations of these invariants. In the CPMSSM 
twelve independent invariant combinations can be defined,   
$$
arg(\mu  M_a  {m_3^2}^{*}) \quad, \quad arg(\mu  A_i {m_3^2}^{*}) \quad \quad .
$$ 
Any other invariant combination is expressed in terms of these. 
%%%%%%%%%%%%%%%%
\begin{table}
\begin{center}
\begin{tabular}[H!]{ccc} \hline 
 Parameter  &   \quad \quad  ${U(1)}_{PQ}$ - charge &  \quad \quad ${U(1)}_{R-PQ}$ - charge  \\  
\hline\hline
$\mu  $    & $-2$ & $2$ \\ 
$h_{t,b,\tau} ... $    & 0& 0 \\ 
$M_a$  & $  0 $ & $-2$ \\ 
$m_3^2$   & $-2$ & 0 \\ 
$A_{t,b,\tau} ...$   & 0 & $-2$ \\ 
$m^2_{\tilde{q},\tilde{l}}$   &  0& 0 \\ 
\hline
\end{tabular}
\caption{Parameters and their $PQ$, $R-PQ$ charges}
\label{tablechar}
\end{center}
\end{table}
%%%%%%%%%%%%%%%%%%%%%

As already discussed the Yukawa couplings can be taken real, positive or negative 
{\footnote{ 
In our convention  
%$ \; W \;=\; \mu H_1 \epsilon H_2 
%  + h_t Q U_c \epsilon H_2 + h_b Q D_c \epsilon  H_1 + h_{\tau} L E_c \epsilon H_1 \;$ 
the VEVs of the Higgses are $\vev{H_1}=v_1$, $\vev{H_2}=\exp (i \theta) v_2$, with 
$v_{1,2}$ real, and trhe superpotential, see \ref{sup}, is such that  
the  running masses for the bottom and tau are $m_b = -h_b v_1 $ and  
$m_{\tau}= -h_{\tau} v_1 $.  We assume $h_b, h_{\tau} < 0$ so that 
these are positive. Since  $\vev{H_2}$ is complex 
the top mass term is $\;h_t v_2 \; e^{\;i\;\theta} \;t t_c + (h.c.) \;$.  
The phase $\; \theta$ can be absorbed by a chiral rotation resulting to a positive   
top  mass $m_t \;=\; h_t v_2\;$ if $h_t > 0$. 
We shall adhere to this convention in the following. 
This chiral rotation modifies other couplings, like 
 gluino-top-stop for instance , which will appear to be $\theta$ dependent. }}.
%$\; (\slash{p}- m_t)$ 
%%%%%%%%%%%%%%%%%%% 
Moreover if they are real at one scale they remain real at all scales. 
The reason is that their RGEs are of the generic form $dh/dlnQ=S\;h$, with $S$ real. 
The slepton and squark soft masses can be also taken real since their 
imaginary parts are completely decoupled from the theory. In general 
the Lagrangian part involving the scalar soft masses has the following  form,
%%%%%%%%%
\bear
{\cal L}_{soft}= - {\frac{1}{2}} \;m_{ij}^{2}\; s_i^* s_j\;+\;(h.c)
\label{softmass}
\eear
%%%%%%%%%%%%%
where summation over the squark, slepton and Higgs fields, denoted by $s_i$, is understood. 
The RGEs of $m^2_{ij}$ at two-loop order can be found
 in~\cite{Martin:1993zk, Yamada:1994id}.
In the MFV scenario the matrix $m_{ij}$ is diagonal and therefore only the real parts of 
the diagonal elements $m_{ii}$ participate  in Eq.~(\ref{softmass})
 and affect physical quantities. 
Their imaginary parts are decoupled. 
Thus, only $Re(m^2_{ii})$ matter whose signs can be either $(+)$ or $(-)$. 
In most of the cases the $(-)$ case leads to potentials breaking colour and/or
 lepton number, or being unstable, and therefore these cases are not phenomenologically 
interesting.  

In order to choose a particular basis to work with, one can exploit the PQ symmetry to 
make $m_3^2$ real. In this basis the invariants under the remaining ${U(1)}_{R-PQ}$ 
symmetry are $ arg(\mu  M_a ) $ and $ arg(\mu  A_i )$. Furthermore, by use of 
${U(1)}_{R-PQ}$ one of the phases of $\mu, Ma, A_i $ can be rotated away and no 
further rotations are allowed. 
Which one is rotated away is a matter of choice.  
For instance in mSUGRA, in the presence of CP-violation and with universal boundary conditions, one usually 
chooses to eliminate the phase of common gaugino mass, $M_{1/2}$, at the unification 
scale and two phases remain, this of $\mu$ and that of the common trilinear 
coupling $A_0$. Since it is customary in mSUGRA to work in the basis in 
which $\mu, A_0$ are complex it is advisable that we work 
in a basis where the phase of $\mu$ is not rotated away either, offering 
the opportunity of a direct comparison of CPMSSM with mSUGRA models.  Also since in the 
EDM cancellation mechanism, to be described later, we implement rotations of 
the phases of $M_1$ and $M_3$, in order to obtain values for the EDMs of 
electron and neutron much smaller than their experimental bounds, we had 
better not rotate away these two phases either.  

One should take care of the fact that, in general, phases run with the energy 
scale. Thus, if a phase is  set to zero  at some energy scale, it may reappear
 at some other scale due to the RGE running. Exception to that are the Yukawa 
couplings and the $\mu$ parameter whose RGEs are multiplicative, by real 
functions, at any loop order. 
The RGEs of the soft gaugino mass parameters are multiplicative at one-loop, 
but not at the two-loop 
order, while those of the trilinear couplings are not multiplicative already at 
one-loop order. Therefore, more phases are expected to be generated through the 
RGE running even if at some scale they are vanishing. In  mSUGRA like models, 
for instance, three phases are generated for the $M_{1,2,3}$ gaugino masses at 
low energy scales, due to the two-loop running, even if some of those are set
 to zero at the unification scale. We shall discuss this issue in detail later on.  

Except the phases associated with the parameters mentioned above, one further phase
 emerges, through loop effects. This is the misalignment angle of the Higgs VEVs 
which is present even if $m_3^2$ is chosen real. To maintain both Higgs VEVs real 
at one-loop an additional rotation of the Higgs fields should be performed, 
but then this relative phase moves someplaces else affecting other parameters. 
In our approach we take $m_3^2$ real at the minimization scale $\; Q_{\mathrm {min}} \;$, 
usually taken to be the average stop mass, by an appropriate PQ rotation. 
The reality of $m_3^2$ simplifies the solution of the minimization conditions 
as we shall discuss. 
In addition one can exploit the $U_Y(1)$ gauge symmetry of the Lagrangian to 
redefine fields in such a way that the VEV of $H_1$ is real. 
Thus, one has $\vev{H_1}=v_1$ and $\vev{H_2}=\exp(i\theta) \; v_2$,  
with $v_{1,2}$ both real. In general they can be taken both complex, 
$\vev{H_i}=\exp(i\chi_i) \; v_i, \; i=1,2$, but only the combination 
$\chi_1+\chi_2=\theta$ is observable. 
In the following we shall work in the basis in which $\vev{H_1}$ is real.  
In this basis it is more appropriate to incorporate $\theta$ with 
the phase $\phi_\mu$ of $\mu$ and use the combinations 
$arg(\mu M_a \exp(i \theta) )=\phi_\mu+\xi_a+\theta$, and  
$arg(\mu A_i \exp(i \theta) )=\phi_\mu+\phi_{A_i}+\theta$  
to express physical observables. 
The reason of doing that is that chargino, 
neutralino and sfermion masses, as well as the EDMs of fermions, 
the quark chromoelectric moments, and the dimension-6 Weinberg 
operator, depend on the combination 
$\phi_\mu + \theta$ rather on $\phi_\mu$ alone \cite{EDMcancel2b}.

In our treatment we take $m_3^2$ and $\vev{H_1}$ real at the 
minimization scale, as we discussed earlier, but we do not implement a further ${U(1)}_{R-PQ}$ transformation to rotate away one of the remaining phases at $\; Q_{\mathrm {min}}$. 
The reason is that even if any of these is rotated out at $\; Q_{\mathrm {min}} $, 
it will reappear  at the unification scale $M_{GUT}$, with the exception of the phase of $\mu$, due to the RGE running.  Therefore we found it more convenient to deal with all thirteen phases, 
those of $\mu, M_a, A_i$ at $M_{GUT}$. The phase of $m_3^2$ at $M_{GUT}$ 
and the misalignment angle of the VEVs, $\theta$, are calculable and not free parameters. 
Although legitimate, in this procedure different choices for the input phases at $M_{GUT}$ 
 may correspond to the same physical situation. Therefore for given SUSY inputs we compare 
the values of $arg(\mu M_a \exp ( i \theta) )$ and 
$arg(\mu A_i \exp ( i \theta) )$ at $\; Q_{\mathrm {min}} \;$ and $\theta$ after 
each run. If they coincide they correspond to the same  physical situation and 
should not be double counted. 
%%%%%%%%%%%%%%%%

Supersymmetric CP-violation has important phenomenological implications and in 
conjunction with other cosmological and experimental constraints deserves further 
detailed study. For a review see \cite{Chung:2003fi} and references therein. 
The presence CP-violating phases affects the cosmological predictions for the 
neutralino relics which are the leading candidates for CDM. They have large 
impact on the bottom mass corrections, $\Delta m_b$, which in turn affects 
drastically the predicted neutralino relic density 
$\Omega_{\tilde \chi} h_0^2$. This effect is more enhanced for large 
values of $\tan \beta$, and for small  trilinear scalar couplings the 
important phases are those of the parameter $\mu$ and the phase $\xi_3$ 
of the gaugino mass $M_3$, \cite{Gomez:2004ef,Gomez:2004ek,Gomez:2004yv}. 
For large values of the trilinear couplings, at low energy scales, their 
phases play a significant role also. The appearance of phases greatly affects 
the masses of the Higgs bosons as well, and thus they have a large impact on the 
DM predictions for neutralino annihilations near a Higgs resonance region, 
which is one of the regions favoured by the cosmological data . 
In  \cite{Gomez:2005nr}  particular extensions of the mSUGRA, 
where the magnitudes of the soft breaking parameters are universal but their 
phases are different in general, were found to be consistent with EDMs, WMAP3 data, 
$b \goes s \gamma$ and $b - \tau$ Yukawa  unification in regions of the parameter 
space in which the phases $ \phi_{\mu}, \xi_3$ have large values. In these 
considerations the cancellation mechanism among the various contributions to 
EDMs was invoked \cite{EDMcancel1,EDMcancel2,EDMcancel2b,EDMcancel3,EDMcancelstring2}, which proved  to be a powerful tool to comply with EDM constraints, in the low $m_0, M_{1/2}$ 
regime, relaxing the stringent constraints imposed on the CP-violating phases 
although its validity and naturalness have been questioned in other works \cite{Hg199}. 

In this work we shall refine the analysis of the CP-violating models, 
focusing on mSUGRA-type models in which all phases are opened up, 
but magnitudes of the SUSY parameters are universal at the unification scale. 
These models are a subclass of the CPMSSM, and their phenomenology has been studied. 
However the subtleties associated with the two-loop running of all phases involved,
 in conjunction with the delicate treatment of the misalignment angle of the Higgs VEVs, 
which arises from the one-loop effective potential, have not been fully treated. 
The effects arising from such a consideration are important for EDMs and relic 
densities as we shall discuss.

\section{CP-violation in the top-down approach}
In the extended class of supersymmetric models discussed in the previous section, 
and in order to 
obtain the mass spectrum and phases at low scales one has to run seventy eight (78) 
RGEs for the real and imaginary parts of all quantities involved, including those 
of the six trilinear scalar couplings of the first two generations which are 
important for the study of the EDMs of the light fermions since they are 
affected by the gaugino masses and the trilinear couplings of the third generation 
species. Due to this dependence non-vanishing phases for these trilinear couplings
 can be developed, even if they are absent at the unification scale, affecting the
 EDMs of the light leptons and quarks  with important phenomenological consequences. 
  
The RGEs for the gauge couplings are certainly real and the Yukawa couplings 
can be taken real as explained in the previous section. The imaginary parts 
of the soft squark, slepton and Higgs masses run with the energy scale but as stated in
 the previous chapter have no effect on physics. Any choice for them leads to 
the same physical results and for convenience we take them zero at the unification 
scale. The RGEs, up to two-loop order, can be found in \cite{Martin:1993zk} 
and \cite{Yamada:1994id} and have been adapted to our own notation (see Appendix). 
 These are run from a GUT scale $M_{GUT}$ defined to be the point at
 which the gauge couplings $\alpha_{1,2}$ unify. We do not enforce unification of
 these with the strong coupling constant although it is left as an option in 
our numerical code. In this procedure we observe radiative electroweak symmetry
 breaking conditions which are altered in the presence of  CP-violating sources. 
The Yukawa couplings of the first two generations have little effect and can 
be neglected from the remaining  RGEs in the approximation that the third generation 
Yukawa's dominate.  

In our analysis we follow a top-down approach with input values for the magnitudes 
and the phases of the soft masses  and trilinear couplings at the GUT scale. 
The reasoning behind this relies on the fact that these are not known at low 
energies but they are rather determined from the fundamental underlying theory, 
which describes physics at GUT or Planckian energies. 

As explained in the previous section we shall work in the basis in which by 
appropriate $U(1)_{PQ}$ symmetry 
the Higgs mixing parameter $m_3^2$ is real at the minimization scale $Q_{min}$
 which we choose to be the average stop masses 
scale $Q_{\tilde t}= { ({m_{{\tilde t}_1} m_{{\tilde t}_2}})}^{1/2}$. 
The neutral Higgses develop VEVs  along the directions
$ \vev{H_1}= v_1 \;,\; \vev{H_2}=v_2 \;e^{i \theta}$ and it is 
convenient to shift the neutral Higgs components as 
%%%%%%%%%
\beq
H_1=v_1+\frac{R_1+i I_1}{\sqrt 2} \;, \; H_2=e^{i \theta}\;\left( \; v_2
+\frac{R_2+i I_2}{\sqrt 2} \; \right).
\label{vacua}
\eeq
%%%%%%%%%%%%%%
As in the CP-conserving case we write 
$v_1 \equiv \frac{v}{\sqrt 2} \cos \beta \;,\; v_2 \equiv \frac{v}{\sqrt 2} \sin \beta $ 
by defining the angle $\beta$. 
The misalignment angle $\theta$ is determined by minimizing the scalar  potential 
and it is vanishing at the tree level. The minimization conditions are then given by 
%%%%%%%%%%%%%%%%%%
\beq
\begin{aligned}
&\frac{1}{2} \; (\; M_Z^2 + \Pi_{ZZ}\;)\;=\;
\frac{ \bar{m}_1^2 \;-\; \bar{m}_2^2 \; {\tan \beta}^2 \; {\cos^2{\theta}} }
{( \;{\tan \beta}^2 - 1 \;)\; (\; c_\beta^2 \;+ \;s_\beta^2 \;{\cos^2{\theta}}\;)}   \\
&\sin{2 \beta} \quad \;=\; 
- \left( \frac{c_\beta^2 \;+\;s_\beta^2 \; {\cos^2{\theta}} }{\cos \theta} \right) \; 
\frac{2 \; m_3^2}{\bar{m}_1^2 \;+\;\bar{m}_2^2 }
 \\
&m_3^2 \; \sin \theta \;=\;\frac{1}{\sqrt 2\; v_2} \; \frac{\partial \Delta V}{\partial I_1}
\; \; .
\label{minima}
\end{aligned}
\eeq
%%%%%%%%%
In the equations 
above $c_\beta \equiv \cos \beta, s_\beta \equiv \sin \beta$, $\Delta V$ 
is the loop corrected scalar potential, and 
%%%%%%%%%
$$
\bar{m}_i^2 \equiv {m}_i^2 \;+\; \frac{\partial \Delta V}{\partial {(Re H_i^0)}^2} \; \;.
$$
%%%%%%%%%
In the first of Eqs~\ref{minima}, ${M}_Z$ is the physical (pole)  Z-boson mass 
and $\Pi_{ZZ}(k)$ is the 
transverse $Z-$ boson propagator correction at $k^2=M_Z^2$. Its inclusion is
 important for a correct numerical treatment. 
Note that the expression on the l.h.s. in the first of Eq. \ref{minima} defines the squared 
of the running  $Z-$ boson mass, 
$ {\hat{M}}_Z^2= M_Z^2 + \Pi_{ZZ}$, which is used to derive the relations 
of $v_{1,2}$ with the other quantities.  
When the CP-violating phases are switched off the third of the equations 
above yields a vanishing misalignment angle 
$\theta$, since its r.h.s. vanishes at the minimization point $I_i, R_i =0$. 
In this case the first two of Eqs. \ref{minima} receive their well-known expressions valid 
in the CP-conserving case.  
In our treatment all one-loop corrections to the effective potential $\Delta V$ have 
been calculated allowing for CP-violating sources in the field dependent masses 
of all SUSY sectors involved, i.e.  sfermions, charginos, neutralinos and Higgses.

In our numerical approach we have $\tan \beta$ as input and the value of $m_3^2$ 
at the minimization scale is determined by the second of Eqs. \ref{minima}. 
The magnitude of $\mu$ parameter is output determined by the first of Eqs. \ref{minima}. 
Recall that $m_i^2= {|\mu|}^2 \;+\; m_{H_i}^2$, $i=1,2$, with $m_{H_i}$ denoting the 
soft Higgs masses. In the presence of CP-violating sources the phase $\theta$ is 
non-vanishing,  because of loop corrections to the effective potential, and it is 
determined from the third of Eqs. \ref{minima} ( see \cite{Demir:1999hj,Demir:1999zb}). 
Thus, it is expected to be small but its impact on the electric 
dipole moments of known species may be sizable. Since this is a one-loop 
quantity one can consistently solve Eq. \ref{minima} by putting $\theta=0$ within $\Delta V$.  

The fields $R_{1,2}, I_{1,2}$ in Eq. \ref{vacua} are not mass eigenstates. A linear combination of $I_{1,2}$, 
namely $ I^{\prime }_2= -c_\beta I_1 + s_\beta I_2$ is the Goldstone mode and the 
orthogonal to it,  
$ I^{\prime }_1= s_\beta I_1 + c_\beta I_2$, gets mixed with $R_{1,2}$ through 
a $3 \times 3$ mass matrix. When CP violating effects are absent this matrix 
does not allow mixing of 
$ I^{\prime }_1$  with $R_{1,2}$. In this case  $ I^{\prime }_1$  is the pseudoscalar, 
CP-odd, mass eigenstate.  The other modes $R_{1,2}$ do get mixed and they must be 
rotated to yield the heavy and light CP-even mass eigenstates . 

The corrections to the masses of the third generation, through which the 
corresponding Yukawa couplings are read,  are very important and affect the 
numerical treatment. The presence of CP violating sources affects these 
corrections  substantially. The supersymmetric corrections to the  bottom 
mass are sizeable for large values of $\tan \beta$ and  should be duly taken 
into account in the analysis. These give rise to large corrections to the 
bottom Yukawa coupling \cite{bottomcor,Carena:1999py,Ibrahim:2003ca} given by 
%%%%%%%%%%%%%%%
\beq
|\; {\hat h}_b \;| \;=\; \frac{{\hat m}^{SM}_b(M_Z)}{v_1} \;{(1+\Delta_{SUSY}^b)}^{-1}  \; \;.
\label{hbot}
\eeq
%%%%%%%%%%%%%%% 
Throughout with  hat we denote quantities  in the $\overline{DR}$ scheme. 
In the equation above the SUSY corrections are denoted by $\Delta_{SUSY}^b$ 
and they are resumed according to the scheme presented 
in \cite{Carena:1999py}. 
%%%%%%%%%%%%%
In this equation ${\hat m}^{SM}_b(M_Z)$ is the Standard Model $\overline{DR}$ 
value of the bottom running mass 
at the scale $M_Z$. Its value is calculated by running the $SU_c(3) \times U_{em}(1)$ 
RGEs  
%%%%%%%%%%
{\footnote{ In our treatment 
we use 3-loop QCD and two-loop QED RGEs for the strong and electric couplings and 
two-loop for the running 
masses of the bottom and tau. Four-loop, ${\cal{O}} ({\alpha}_s^4)$, 
QCD contributions to beta functions and quark anomalous dimensions are 
available \cite{Chetyrkin:1997dh} but affect little our analysis.
}} 
%%%%%%%%%%%%%%%
for masses and couplings in the $\overline{MS}$ scheme, from 
the bottom mass  ${\hat m}^{SM}_b(m_b)=4.25 \pm 0.15  \GeV$ \cite{Martinelli:1998vt}, 
as determined in lattice calculations, 
up to the scale $M_Z$. Its $\overline{MS}$ value at $M_Z$ is
 subsequently converted to its $\overline{DR}$ value using 
well-known expressions. Therefore from Eq. (\ref{hbot}) the 
value of ${\hat h}_b$ can be extracted which is needed to run 
the RGEs from $M_Z$ to the GUT scale. The corrections involved within 
$\Delta_{SUSY}^b$ are very important and are discussed below.

The leading supersymmetric QCD, sbottom-gluino, and Electroweak (EW), 
stop-charginos, contributions  to $\Delta_{SUSY}^b$ are given by 
%%%%%%%%
\bea
\Delta_{SUSY}^b &=& \frac{2 \alpha_s}{3 \pi}\; M_{\tilde g}\;
( \;|\mu | \tan \beta \cos (\xi_3+\phi_\mu +\theta) + |\mu || A_b | 
   \cos(\xi_3- \phi_b\;)  \;) 
\;G( {\tilde b}_1,{\tilde b}_2,{M_{\tilde g}})  
\nonumber \\
\;&&-\; 
\frac{h_t^2}{16 \pi^2}\; | \mu | \;(\;| \mu | + | A_t | \tan \beta \;
              \cos( \phi_\mu +\phi_t +\theta ))\;
G( {\tilde t}_1,{\tilde t}_2,| \mu |) \;. 
\label{deltab}
\eea
%%%%%%%%%%
$|A_{b,t}|,| \mu |$ are the magnitudes of $A_{b,t}, \mu $  and $\phi_{b,t}, \phi_\mu$ 
their phases. $\theta$ is the misalignment angle between the Higgs VEVs and tilded 
quantities denote sbottom, stop masses
%%%%%%%%%%%%%%%%%%%%%%%%%%%%%
{\footnote{The function $G(a,b,c)$ is identical to $I(a,b,c)$ 
defined in Eq. 7 in~\cite{Carena:1999py}.}}.
%%%%%%%%%%%%%%%%%%%%%%%%%%%
In this expression we have neglected electroweak (EW) mixings of the 
stops, and also sbottoms, and the mass of the chargino is put to $\mu$. 
More refined expression which take into account the mixings can be found 
in ref \cite{Ibrahim:2003ca}. 
An analogous treatment holds for the corrections to the tau mass as well. 
However their effect is small due to the absence of QCD corrections at the 
one-loop level. When the magnitudes of the trilinear couplings involved in 
Eq.~\ref{deltab} are small and the angle $\theta$, which is anyway small, 
are neglected then this equation receives a much simpler form
%%%%%%%%%%%%%%%%%%%%%
\bea
\Delta_{SUSY}^b &=& \frac{2 \alpha_s}{3 \pi}\; M_{\tilde g} \;|\mu | \; 
\tan \beta \; \cos( \xi_3 + \phi_\mu )
\; G( {\tilde b}_1,{\tilde b}_2,{M_{\tilde g}})  \nonumber \\
\;&&-\; 
\frac{h_t^2}{16 \pi^2}\; {| \mu |}^2\;
G( {\tilde t}_1,{\tilde t}_2,| \mu |) \;. 
\label{deltab2}
\eea
%%%%%%%%%%%%%%%
The functions $G$ appearing in Eq. \ref{deltab2} are positive in a 
large portion of the parametric space and 
therefore if $\cos( \xi_3 + \phi_\mu ) $ is negative the corrections to 
$\Delta_{SUSY}^b$ may turn out to be negative and sizable, in the large $\tan \beta$ regime. 
In this case the bottom Yukawa coupling of Eq. \ref{hbot} gets large and a
 Landau pole may develop. Therefore in the top-down approach, depending on the inputs, 
the approach to the large $\tan \beta$ regime is not guaranteed. This observation 
is important for the mechanism of neutralino annihilation through a Higgs resonance, 
which opens for large values of $\tan \beta$. 
The importance of this will be discussed later. 

The Yukawa coupling of the top quark is large and it is determined from the 
experimentally measured top pole mass in the following way. For the top quark 
the relation between its pole and running mass, including  the dominant QCD and 
the supersymmetric gluino-stop corrections is given by
%%%
\bea
M_t^{pole} \;&=&\;m_t(M_t^{pole})(1+\Delta_{SUSY}^t)/(1-\Delta_{QCD}) \; .
\label{qcdt}
\eea 
%%%
The pole mass is scheme independent so that the r.h.s can be calculated in either
 $\overline{MS}$ or $\overline{DR}$ scheme. 
We prefer to employ the first scheme since the QCD corrections $\Delta_{QCD}$
 have a simple form which at two-loops is given by 
%%%
\bea
\Delta_{QCD}=\frac{4}{3 \pi} \alpha_s + 1.11 {\alpha_s}^2 \; \; .
\label{qcdt2}
\eea
%%%
In Eq. \ref{qcdt2} the strong coupling constant is meant at $M_t^{pole}$ and its 
running from any lower scale to the pole top mass is done using the five quark flavour RGEs.
 Note that in Eq. \ref{qcdt} the QCD corrections have been resummed. The $\overline{MS}$ 
strong coupling $\alpha_s$ appearing in the above expressions is different from the 
corresponding $\overline{DR}$ strong coupling usually denoted by $\alpha_3$. 
The relation between these two will be given in the sequel. 

The gluino-stop corrections appearing in Eq. \ref{qcdt}  
are given by  
%%%%%%%%%%%
\bea
\Delta_{SUSY}^t \;&=& \frac{\alpha_s}{3 \pi}\; 
[ \; - B_1(0,M_{\tilde g},m_{\;{\tilde t}_1}) - 
B_1(0,M_{\tilde g},m_{\;{\tilde t}_2}) \nonumber \\
&-& 
\frac{M_{\tilde g}}{{ m}_t(M_t^{pole})} \; \sin (2 \theta_t) \;
\cos \xi \; ( \;B_0(0,M_{\tilde g},m_{\;{\tilde t}_1})- B_0(0,M_{\tilde g},
     m_{\;{\tilde t}_2})\;)\; ]\; 
 \nonumber
\eea
%%%%%%%%%
In this $\theta_t$ is the angle diagonalizing the stop mass matrix, 
$\xi \equiv \xi_3+\phi+\theta$ and 
$ m_{\;{\tilde t}_{1,2}} $ are the stop masses 
%%%
\footnote{
The diagonalizing matrix $K$ is defined by 
$K M^2 K^{\dagger}=diag \;( m_{\;{\tilde t}_1}^2,m_{\;{\tilde t}_2}^2 )\; $ 
with the matrix elements of $K$ given by $K_{11}=K_{22}=\cos \theta_t$, 
$\;K_{12}=-K_{21}^{*}=e^{i {\phi}} \;\sin \theta_t$. 
The definition of the matrix $K$ we adopt is consistent with $K$ tending to
 the unit matrix if the 
off-diagonal elements of the squark mass matrix are switched off. 
Thus, the subscripts in 
$m_{\;{\tilde t}_{1,2}}$ do not specify the ordering of their heaviness.}.
%%%%%%%%%%%%%%%%
This generalizes the results of ref. \cite{Pierce:1996zz} in the case that the 
supersymmetric parameters are complex. 
The functions $B_{0,1}$ are defined as in \cite{Pierce:1996zz}. 
A minus sign difference with the results of that reference, 
occurring in the case of the absence of CP violations, $\xi=0$, 
is due to the slightly difference notation used here.
Note that these corrections have the same form in both $\overline{MS}$ 
and $\overline{DR}$ schemes since in the gluino-stop loop scalar particles 
are exchanged and no traces of gamma matrices are involved.
 Their difference in the two schemes is therefore small, two-loop, due to the fact 
that the couplings and running masses appearing already differ 
at one-loop in the two schemes. 
More refined relations including the subdominant EW
corrections are given in  \cite{Ibrahim:2003ca}.  
The EW supersymmetric corrections are however small and  
they correct the approximate result by less than 1\% which becomes 
even less when the stop masses are of the order of 1 TeV. 
%%%
Therefore the $\overline{MS}$ value for the top Yukawa coupling is given by
%%%%
\bea
h^{\overline{MS}}_t (M_t^{pole}) \;=\; \frac{M_t^{pole}}{v_2} \; 
        \frac{1- \Delta_{QCD}}{1+\Delta^t_{SUSY}} \; {\Bigg\vert}_{\overline{MS}}
\eea
%%%
and its $\overline{DR}$ value needed to run the corresponding RGEs is 
                    provided by the usual conversion formula \cite{Martin:1993yx}
%%%%
\bea
h^{\overline{MS}}_t (M_t^{pole}) \;=\; h^{\overline{DR}}_t (M_t^{pole}) \; 
( \; 1 + \frac{\alpha_3}{8 \pi}+ \frac{\alpha_2}{16 \pi}+\frac{3 \alpha_1}{80 \pi} \;) \; \; .
\eea
%%%%

Besides the input Yukawa couplings, we need the values of the gauge couplings 
at the unification scale $M_{GUT}$ defined to be at the point where the gauge 
couplings $\alpha_{1,2}$ meet. These are determined by their values at the 
scale $M_Z$ in terms of the fine structure constant $\alpha_0$, the value of
 the Fermi coupling constant $G_F$ and the physical Z-boson mass $M_Z$. 
These are related to  $\alpha_{1,2}(M_Z)$ in the way prescribed in~\cite{Pierce:1996zz}. 
These are run up by 
two-loop RGEs and the unification scale  $M_{GUT}$ is determined at the 
point where these intercept. Naive gauge coupling unification would entail 
to putting the strong coupling $\alpha_3$ equal to $\alpha_{1,2}$ at $M_{GUT}$. 
Instead of doing this we follow the alternative, followed in \cite{Pierce:1996zz},
 according to which the $\overline{MS}$ value of the strong coupling, denoted 
by $\alpha_s$, coincides with the one experimentally measured. Its relation 
to $\alpha_3$ at the scale $M_Z$ is given by 
%%%
$$
\alpha_s(M_Z)=\alpha_3(M_Z) /(1-\Delta \alpha_3)
$$ 
%%%%
where $\Delta \alpha_3$ includes supersymmetric threshold corrections and constants 
associated with passing from $\overline{MS}$ to $\overline{DR}$ scheme. 
This determines $\alpha_3$. 
The Yukawa and gauge couplings determined at lower scales in the way prescribed earlier 
are run up at $M_{GUT}$ to determine their values at the unification scale. This
 is done iteratively until a certain convergence is achieved. In our code the 
Higgsino and Higgs mixing parameters $\mu, m_3^2$ are outputs, and the 
convergence criteria in our numerical code are tailored to monitor these 
parameters in each iteration, which in most of the cases have the slowest 
convergence from the other parameters involved. In determining these we 
use the full one loop effective potential with the leading two-loop QCD 
correction taken into account. The minimization conditions are considered 
at an average stop scale as described earlier and the Higgs mixing parameter 
$m_3^2$ is taken real without loss of generality as we have discussed. 
At any other scale, including the unification, this is certainly complex 
and its phase is extracted from the running of the RGEs from the minimization scale. 

Concerning the mass spectrum of SUSY particles the gluino physical mass is 
overwhelmed by large QCD corrections which affect the numerical analysis and 
should be discussed. These are due to both SM gluon exchanges and 
supersymmetric corrections due to exchanges of fermions and their corresponding 
squarks. The relation between the  physical, $M_{\tilde g}$, and the soft 
gluino mass, $M_3$, is found to be  
%%%%%%%%%%%%%%%%%%
\beq
M_{\tilde g}\;=\; \frac{| M_3(Q) |}
{(\;1-3 \alpha_3(Q) \;(\;5+6 \ln (Q/|M_3(Q)|)- S(Q))/(4 \pi)}
\eeq
%%%%%%%%%%
where $\alpha_3$ is the ${\overline{DR}}$ value of the strong coupling and $S(Q)$ 
the squark contribution given by
%%%%%%%%%%
\beq
S(Q)= \frac{1}{3} \; \sum_{i=1}^3( 2{\tilde q}_i+{\tilde u}_i+{\tilde d}_i)\;.
\eeq
%%%%%%%%%
In $S(Q)$ ${\tilde q}_i$ denotes the contribution of the squark doublets, 
accommodating  the left-handed up and down squarks,  and ${\tilde u}_i,{\tilde d}_i$
 are those of the corresponding right-handed squarks. In both cases the index $i$ 
runs over the colour. 
For ${\tilde q}_i$, and the same holds for ${\tilde u}_i,{\tilde d}_i$, 
%%%%%%%%%%555
\beq
{\tilde q}_i\;=\;- \frac{1}{2} \;\ln(\;\frac{M^2_i}{Q^2}\;)\;+1- \frac{1}{2r} \;
(1+ \frac{{(r-1)}^2}{r} \ln {|r - 1|})\;+\;\frac{1}{2}\;  \theta(r-1)\; \ln r
\eeq
%%%%%%
where $M_i=max( {\tilde m}_i , |M_3|\;) $, $r= {|M_3|}^2/{{\tilde m}_i }^2$. 
${\tilde m}_i $ is the squark soft mass in each case. We have neglected 
EW mixing effects which have little effect on this formula. 
This generalizes the result of  \cite{Pierce:1996zz}, in which a common 
squark mass is used, and it is a handy expression to use avoiding the 
complexities of other calculations. 
More refined two-loop corrections in terms of several two-point functions are presented in 
\cite{Yamada:2005ua,Martin:2005ch}. 
In that reference it is shown that the two-loop QCD corrections are small. 

With the above we end the discussion concerning the treatment of the RGEs by 
giving an outline of the salient features which have a large impact on 
our numerical analysis in the CP-violating case. The boundary conditions
 employed for the couplings at the unification scale were discussed earlier. 
In our code the soft masses and trilinear scalar masses we treat in the most 
general case allowing for  non-universal boundary conditions for their magnitudes
 and their phases, as already discussed, in the approximation of neglecting 
flavour violating interactions in the Lagrangian. 
Although the setup is to cover the most general case in this respect we shall 
only discuss particular cases which are of phenomenological and theoretical 
interest, and focus our attention mainly on models with universal boundary 
conditions for the magnitudes of the SUSY breaking parameters involved.  

\section{Running of the CP-violating phases with the energy scale}
The case of non-vanishing phases has features that need be further discussed when the 
two-loop RGEs are used for their evolution. Their values are inputs  which can be set
 at either the unification scale, $M_{GUT}$, or the average stop scale $Q_{\tilde t}$ 
at which we minimize the one-loop effective potential. In our   treatment  we  work in 
the basis in which the phase of $m_3^2$ is put to zero at $Q_{\tilde t}$ and the 
remaining phases we take as inputs at  $M_{GUT}$, as we have already stated.  
Evidently the phases of all parameters at any other scale are determined by the RGE running. 
The phase of the $\mu$ is very important since it affects the mass spectrum of the 
charginos, neutralinos and  sfermion and it explicitly appears within their 
corresponding mass matrices. Besides, its phase has a large impact on the 
radiative corrections of the bottom and top Yukawa couplings and therefore 
the numerical procedure is very sensitive to its input value. Last, 
but not least, the phase of $\mu$ affects the EDMs of Hg, neutron and electron. 
A large phase for $\mu$ may be in accord with Baryogenesis but its value 
is severely constrained by the experimental EDM bound of the electron in 
mSUGRA models in which only the common gaugino phase and the phase of $\mu$ are present. 

The RGEs of the parameter $\mu$, the gaugino masses and the electron's trilinear 
coupling, which also affect the light fermion  dipole moments, are as follows
%%%%%%%%%
\bea
\frac{d \mu}{d ln Q}\;&=&\; \frac{\mu}{4 \pi} \; ( -3 \alpha_2-\frac{3}{5}  
  \alpha_1 + 3 h_t^2 + 3 h_b^2 + h_{\tau}^2 + {\mathrm{two-loop}}) \nonumber \\
\frac{d M_i}{d ln Q} \;&=&\; \beta_i M_i \;+ \mathrm{two-loop} \,,   
    \quad i=1,2,3  \label{rges}  \nonumber \\ 
\frac{d A_e}{d ln Q}\;&=&\;\frac{1}{4 \pi} \;
 ( - 6 \alpha_2 M_2 - \frac{18}{5} \; \alpha_1 M_1 + 6 h_b^2 A_b 
  + 2 h_{\tau}^2 A_{\tau} ) +{ \mathrm{two-loop}} \quad \quad .
\eea
%%%%
For lack of space we do not present the RGEs of the remaining trilinear 
couplings of the 1st and 2nd  generations and we also avoid presenting explicitly the 
two-loop contributions. The $\beta_i$ in the RGEs for 
the gaugino masses are the one loop beta function for the gauge couplings 
$\alpha_i$. Since we work in the MFV scenario we have neclected generation mixing terms. 
In this approximation the trilinear couplings of $A_e$, as well as the remaining 
trilinear couplings of the 
first two generations, do not have any influence on the remaining parameters 
although they depend on them. In practice this means that one can first solve 
the RGEs for the rest of the parameters and subsequently  determine the trilinear
 couplings of the first two generations.  

The RGEs for the $\mu$ parameter has the form $d \mu / d ln Q = \mu \; S \;$ 
where $S$ is real. Due to this the phase of $\mu$ does not get renormalized 
with the scale. Its value at any scale is the same with its value at $M_{GUT}$ 
independently of the other phases. The phases of the soft gaugino masses exhibit 
a slightly different behaviour. Their one-loop RGEs are multiplicative as in the
 $\mu$ case so that they do not get renormalized at this loop order independently
 of the remaining phases. However this does not hold at the two-loop order. 
This essentially means that one  should expect small renormalization of their 
phases as we run down from the unification to lower scales and vice versa. 
Thus, even if their phases at the unification scale are set to zero, small 
phases are induced at low energies, if the phase of $\mu$ or the trilinear 
couplings are non-vanishing at $M_{GUT}$. Although small this phenomenon may 
have dramatic consequences for the EDMs since even small phases may produce 
large EDMs which are constrained by the data. This effect is more enhanced for
 the phases of the trilinear couplings which are considerably renormalized 
since their RGEs are not multiplicative, already at the one-loop, unlike the 
parameter $\mu$ and the gaugino masses. For instance in the case of $A_e$ 
its one-loop RGE shows a dependence on the gaugino masses and the trilinear 
couplings of bottom and tau, as is obvious from the third of Eqs. \ref{rges}. 
Thus, a non-vanishing phase, for at least one of them, at the unification scale 
may yield a non-vanishing phase at the EW scale for $A_e$ which in
 turn affects the EDM of the electron. The same holds for the remaining 
trilinear couplings. This digression shows how important is to consider the
 running of the phases as we do in our approach.

%%%%%%%%%%%%%%%%
\begin{table}
\begin{center}
$M_{1/2}=900 \;$, $m_0=800\;$, $A_0=500\;$, $m_t=171.4\;$, $m_b=4.25\;$  \\ 
\begin{tabular}[H!]{cc|cccccc} 
\hline \hline
  $\xi_3=2\pi/10$& & &$arg(M_1)$   & $arg(M_2)$  & $arg(A_e)$ & $\theta$ & 
 $10^{26} \times d_e$ \\  
\hline
$\tan \beta=50$& & &-0.0144 &  -0.0198 &  -0.1735 & 0.0465 & +1.0845\\ 
$\tan \beta=10$& & &-0.0149 &  -0.0209 &  -0.0079 & 0.0006 &-0.1654 \\
\hline
\end{tabular}
\caption{The induced phases, at $M_Z$, of $M_{1,2}, A_e$, the misalignment angle $\theta$ 
and the value for the electron EDM when only the gluino phase $\xi_3= \pi/5 $ 
is switched on at the unification scale, for $\tan \beta=50, \;10$ respectively. 
The inputs (in GeV) are shown on the top of Table. $M_{1/2}, m_0, A_0$ are the magnitudes 
of the common gaugino and scalar masses and trilinear couplings at $M_{GUT}$. 
}
\label{tablex3}
\end{center}
\end{table}
%%%%%%%%%%%%%%%%%%%%%

\begin{figure}[t]
\begin{center}
\includegraphics[width=7.7cm]{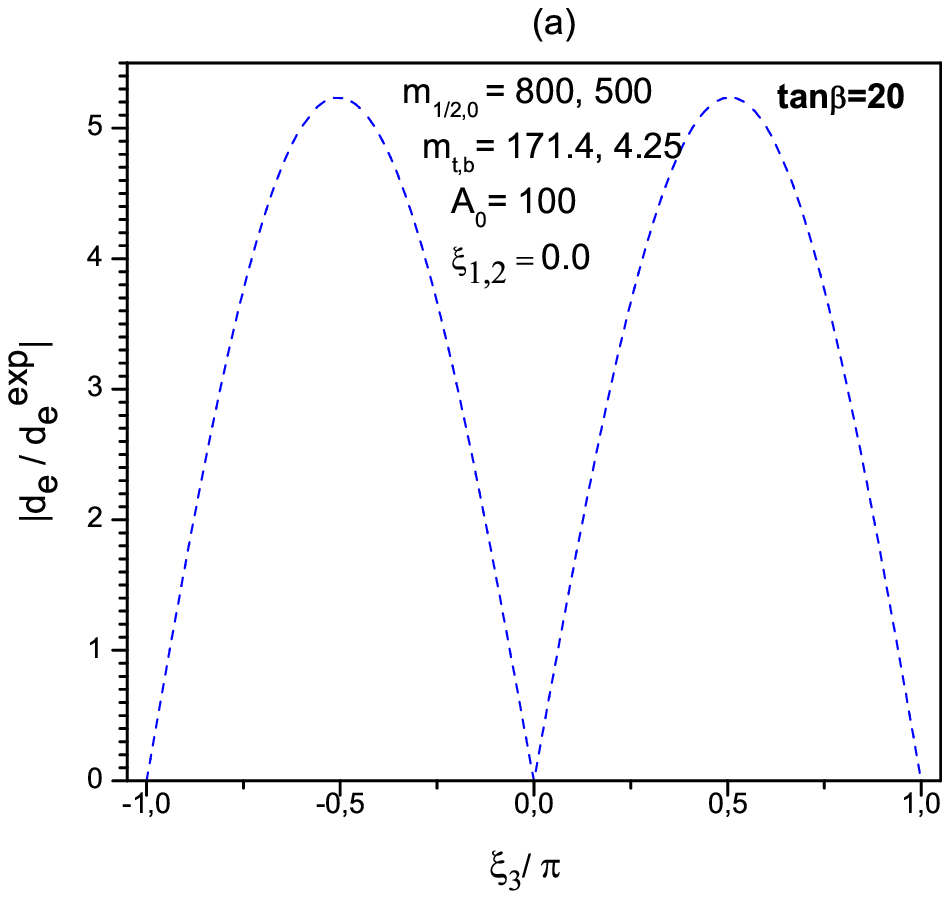}
\hspace*{0.2cm}
\includegraphics[width=7.7cm]{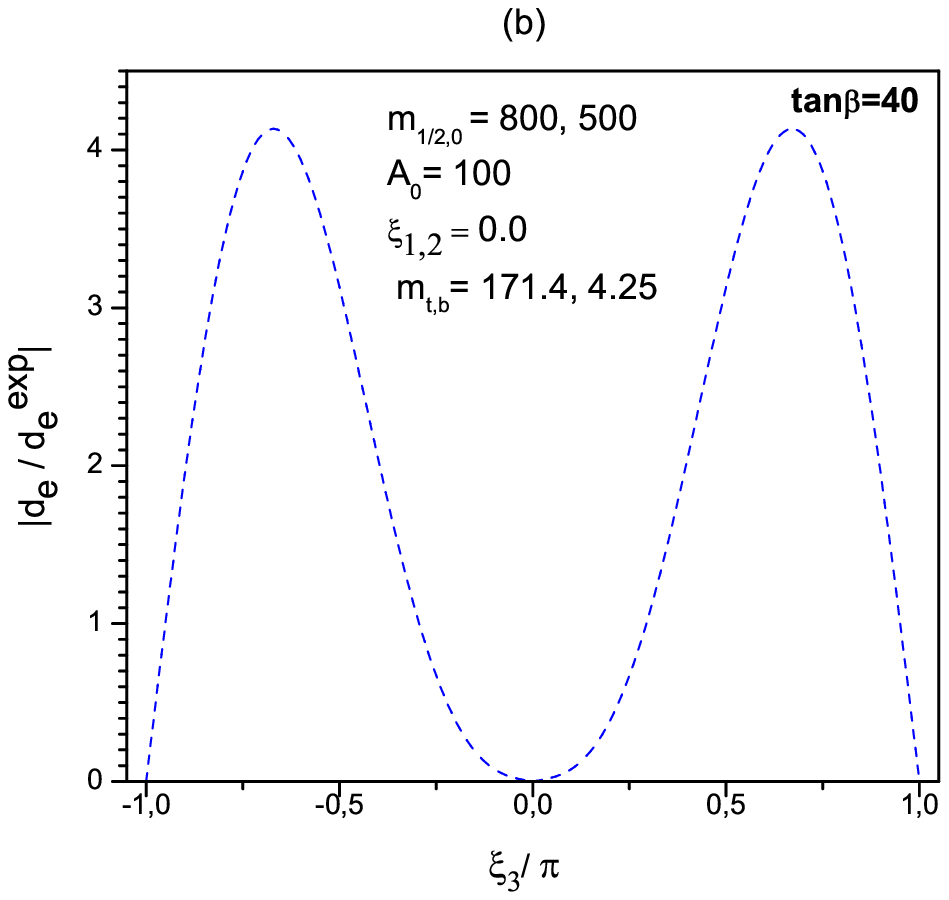}
\end{center}
\caption[]{
The ratio $|d_e/d^{exp}_e|$ of the predicted electron's edm to its 
experimental bound as function of the gluino phase 
$\xi_3$, for two different values of $\tan \beta = 20$ (left panel) and  
$\tan \beta = \;40$ (right panel). The remaining inputs are displayed in the figure.
}
\label{fig1}
\end{figure}

As a preview of the impact of the two-loop RGE corrections to the phases, and consequently 
on EDMs,  consider the 
particularly interesting case arising when the gluino phase $\xi_3$ is the only 
non-vanishing phase at $M_{GUT}$. As is obvious from the above RGEs 
the phases $\xi_1, \xi_2$  of  $M_1,M_2$ as well as this of $A_e$ 
are not affected by $\xi_3$ at one-loop running. 
{\it{
Therefore at this loop order the electron EDM bound does not depend on $\xi_3$.}}
However at the two-loop order the phases for $M_1, M_2, A_e$ do depend on $\xi_3$ 
affecting the EDM, $d_e$, of the electron. In some cases this dependence may have 
important consequences inducing  
corrections to EDMs that are comparable to those induced by the Barr-Zee type 
two-loop contributions which are known to be sizable for large $A$ and $tan \beta > 30$.
If, for instance, $\xi_3$ is the only non-vanishing phase switched on at 
the unification scale the induced phases for $M_1, M_2, A_e$ may result to
values for  electron's EDM  that can even saturate the experimental limits put 
on $d_e$. A typical case is shown in Table \ref{tablex3}, 
for $\tan \beta=50$ and $\tan \beta=10$ respectively. 
The remaining inputs are as displayed in Table.  
Throughout this paper, if not otherwise stated, $M_{1/2}, m_0$ and $ A_0$ will denote the magnitudes of the common  soft gaugino and scalar masses and the common trilinear scalar coupling respectively.  
For $\xi_3 = 2 \pi/10$  at $M_{GUT}$ the phases of $M_{1,2}$ 
at the EW scale are $ \sim 10^{-2}$.  
The phase of $A_e$ is $\phi_{A_e} \sim 10^{-1} ( 10^{-2})$ 
for $\tan \beta = 50 (10)$. For completeness we also display 
the value of the calculated misalignment angle $\theta$ which 
also affects EDMs. It is seen that 
values for $d_e$ which are larger than the experimental limits 
quoted in literature, $|d_e| < 1.6 \times 10^{-27}$, 
are induced for these particular inputs. However this is a generic feature valid 
in a large regions of the $m_0 -M_{1/2}$ plane. 
In Fig.~\ref{fig1} we display the ratio 
$|d_e/d^{exp}_e|$ of the predicted electron's edm to its experimental 
bound as function of the gluino angle $\xi_3$, which is assumed to be the 
only non-vanishing input CP-violating input at $M_{GUT}$, for two 
different values of $\tan \beta = 20, \;40$. The remaining inputs 
are displayed in the figure. This ratio should be less than unity for
 the experimental bound to be observed. A strong dependence on the angle 
$\xi_3$ is seen which is absent at the one-loop RGE running. 
This restricts the allowed $\xi_3$ values to be in the vicinity 
of $0, \; \pm \pi$ for $\tan \beta =20$. Increasing to $\tan \beta = 40$ 
the dependence of $d_e$ is still strong, but slightly milder, and the 
allowed range for $\xi_3$ gets broadened. In this case one observes that 
large $\tan \beta$ values allow for non-trivial gluino phases which are 
welcome since they can be used in order to lower the neutron's edm 
by using the cancellation mechanism as we shall discuss.

%%%%%%%%%%%%%%%%%%%%
%%%%%

The dependence of the $M_1, M_2, A_e$ phases on $\xi_3$ may also affect the 
cancellation mechanism if phases are inputs at the GUT scale. Rotating the 
phase $\xi_1$ till $d_e$ becomes vanishingly small, to comply with its 
experimental bounds, a subsequent rotation of $\xi_3$ at $M_{GUT}$, in 
order to make neutron's EDM small within its experimental limits as 
prescribed in \cite{EDMcancelstring2}, 
will shift the initially found $\xi_1$ invalidating the cancellation 
between neutralino and chargino contributions in $d_e$. Certainly this 
is not the case when this procedure  is implemented with phases given at 
the EW scale. 
Therefore the determination of cancelling phases in the top-bottom approach 
poses difficulties  not encountered in the one-loop RGE running. In addition, 
even if proper phases are found at $M_{GUT}$ by the cancellation mechanism, 
so that electron and neutron EDM become small for some particular SUSY inputs, 
it is difficult to delineate regions by merely rescaling the SUSY parameters 
as prescribed in \cite{EDMcancelstring2}. The reason is that due to their RGE 
running the induced phases at low energies depend on SUSY inputs and a rescaling 
dislocates the values of the low energy phases by little amounts but enough to 
make the cancellation invalid. We are therefore arguing that the cancellation 
mechanism is best suited for a bottom-up approach although it may still be a 
powerful tool to locate regions compatible with EDMs and all other data when 
CP-violating phases are switched on as we shall discuss. 

%%%%%%%%%%%%%%%
\begin{figure}[t]
\begin{center}
\includegraphics[width=7.7cm]{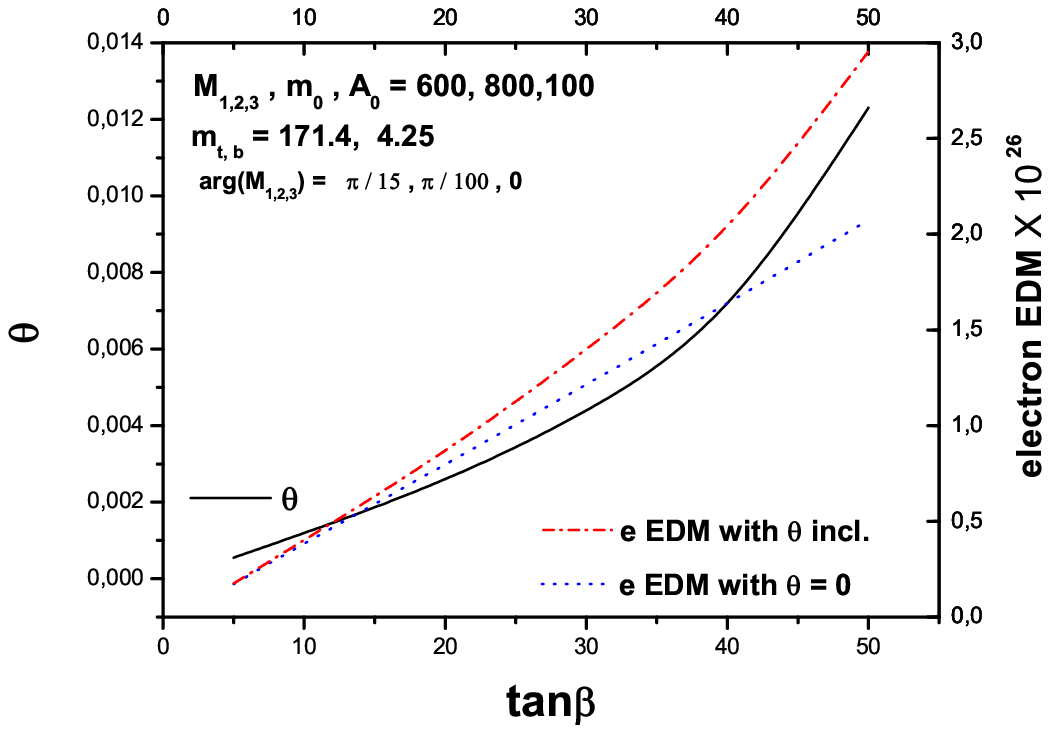}
\hspace*{0.2cm}
\includegraphics[width=7.7cm]{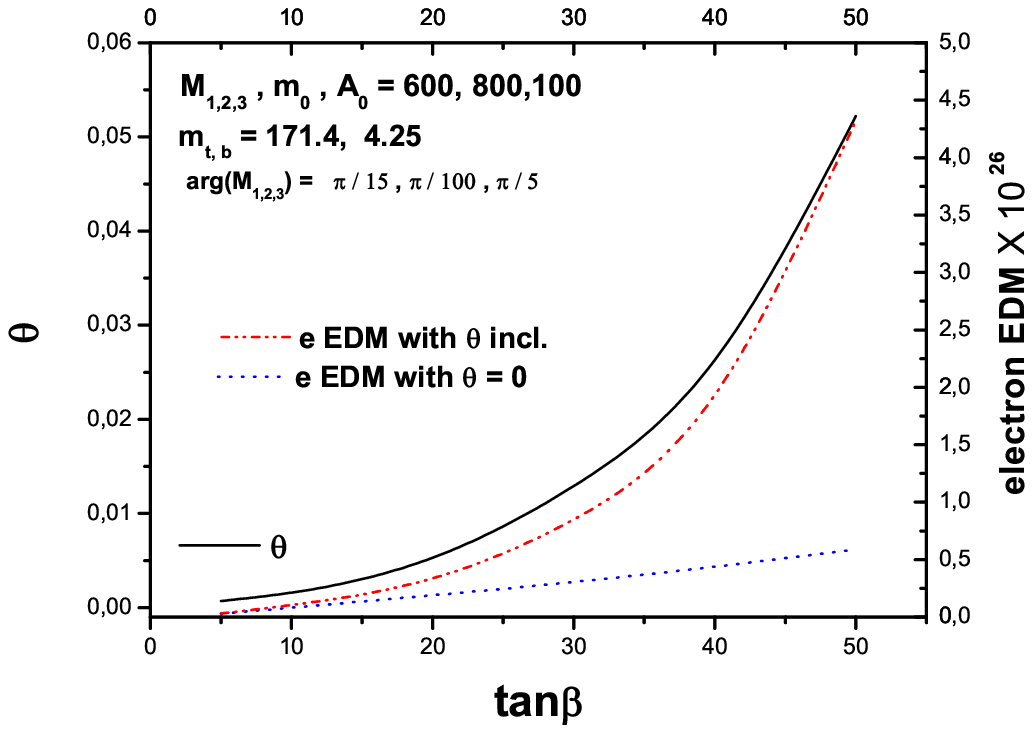}
\end{center}
\caption[]{The misalignment angle $\theta$ (solid line) and the electron 
dipole moments, with $\theta$ included (dashed-double dotted line) and 
with $\theta$ set to zero (dotted line), for the inputs shown on the 
figure. Masses are given in $\GeV$ and $\;M_{1,2,3}, m_0, A_0\;$ refer 
to the magnitudes of the corresponding parameters. On the left (right) 
panel the $M_3$  phase is set to $0 \;(\pi/5)$. 
}
\label{figtheta}
\end{figure}
%%%%%
%%%%%%%%%%%%%%%%%%%%%%%%%
%%%%%%%%%%%%%%%
\begin{figure}[h!]
\begin{center}
\includegraphics[width=10.0cm]{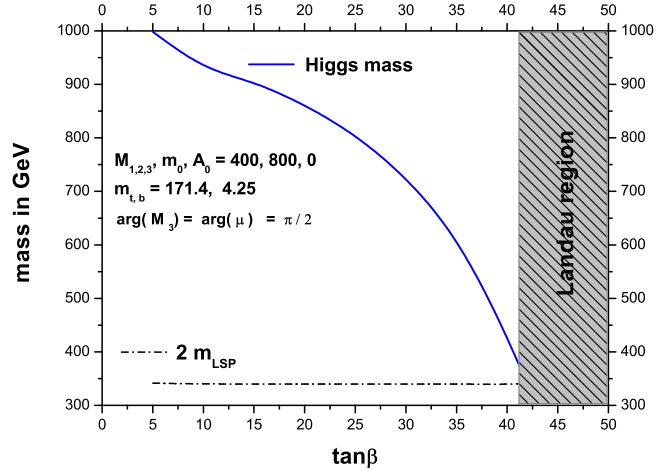}
\end{center}
\caption[]{
One of the heavy neutral Higgs mass (solid line)  and the value of twice the
 LSP neutralino mass 
(dash-dotted line) as functions of $\tan \beta$ for the inputs shown in the figure. Above 
$\tan \beta \simeq 41.0$ a Landau pole is developed. 
}
\label{figLandau}
\end{figure}
%%%%%

%%%%%%%%%%%%%%%

The role of the misalignment angle $\theta$ to the EDMs should not be passed 
unnoticed. This is measurable and cannot be rotated 
away~\cite{Demir:1999hj,Demir:1999zb,Chung:2003fi}. 
Its value enters and affects various physical quantities.
 The neutralino, chargino, squark and slepton masses depend on this 
through the combination $arg(\mu)+\theta$ as has been already stated. 
It also affects the Higgs decays to $b \bar{b}$ 
enhancing the widths for these decays \cite{Demir:1999hj} which has 
important implications for the  cosmologically acceptable 
regions in which LSP pair annihilation takes place near a Higgs resonance. 
 This mechanism depends sensitively on the corresponding widths. 
It has also impact on EDMs especially in the large $\tan \beta$ regime.
 In Fig.~\ref{figtheta}, for some particular inputs, we display the values
 of the angle $\theta$ and the electron dipole moment, $d_e$, with and without 
the inclusion of $\theta$ in its calculation. The angle $\theta$ takes values 
from $\sim 10^{-4} $ to $\sim 10^{-2} $ for  $\tan \beta$ in the region $5 - 50$. 
On the left panel the gluino phase has been taken vanishing at the GUT
 scale and the difference in $d_e$reaches $50 \%$ for  $\tan \beta =50$. 
On the right panel in addition the gluino phase is switched on and the 
difference gets much larger. Therefore the misalignment angle produces 
large effects when $\tan \beta$ is large, which are further augmented 
if in addition the gluino phase is large at the unification scale. 
Since $\theta$ has large impact on EDMs, in particular regions of the 
parameter space, it influences the cancellation mechanism as well,
 in the large $\tan \beta$ region, especially when large values for 
the gluino phase are required to make neutron's EDM small within its experimental bounds. 

As already stated in this section, the phases of the Higgsino parameter 
$\mu$ and that of the soft gluino mass $\xi_3$ affect the analysis a great 
deal. The first, if large, affects the EDM of electron which imposes the
 stringent constraint on the phase $\phi_{\mu}$ of $\mu$, while both 
affect the corrections to the bottom mass especially for large values 
of $tan \beta$ having a large impact on the DM relic density. This is 
clearly seen in Eq. \ref{deltab} or its more simplified form \ref{deltab2}
 where the corrections to bottom mass are encoded in. From Eq. \ref{deltab2} 
we see that if $\phi_\mu, \xi_3$ are such that $\cos(\phi_\mu + \xi_3)< 0 $,  
at low energies, these corrections, depending on inputs,  can be large and negative, 
in the large $\tan \beta$ regime. Therefore  in view of Eq. 
\ref{hbot} they may yield large values for the bottom Yukawa coupling. 
In this case one should be prepared to encounter the appearance of Landau poles and 
the top-down approach can not be handled  perturbatively. Therefore the link between
 low energy and GUT scale physics is questioned in this case. This behaviour imposes a severe obstacle 
when large phases are sought in conjunction with large $\tan \beta$ values which is the
 requirement for the LSP annihilation through a Higgs resonance. 
In Fig.~\ref{figLandau} we present such a situation for the inputs displayed on the figure.
 In this figure we observe the evolution of one of the heavy neutral Higgses mass,
$m_{H_3}$, and the double of the  LSP neutralino mass, $2 \; m_{LSP}$, as functions 
of $\tan \beta$. The Higgs mass tends to $2 \; m_{LSP}$ as $\tan \beta$ increases 
and would eventually catch the $2 \; m_{LSP}$ line signaling approach to a point 
where LSP pair annihilation through a Higgs resonance dominates the relic density. 
However it is shown that this is abruptly stopped due to the development of a Landau 
pole at around 
$\tan \beta \simeq 41$. This effect, not considered in previous analyses, may exclude 
particular points in the parameter space and shrink the allowed funnel regions.   
%%%%%%%%%%%%%%%%%%%%%%%%%%%%%%%%%%%%%%%%%%%%%%%%%%%%%%%%%%%%%%%%%%%%%
\section{Cosmologically and EDMs allowed domains}
\begin{figure}[t]
\begin{center}
\includegraphics[width=7.7cm]{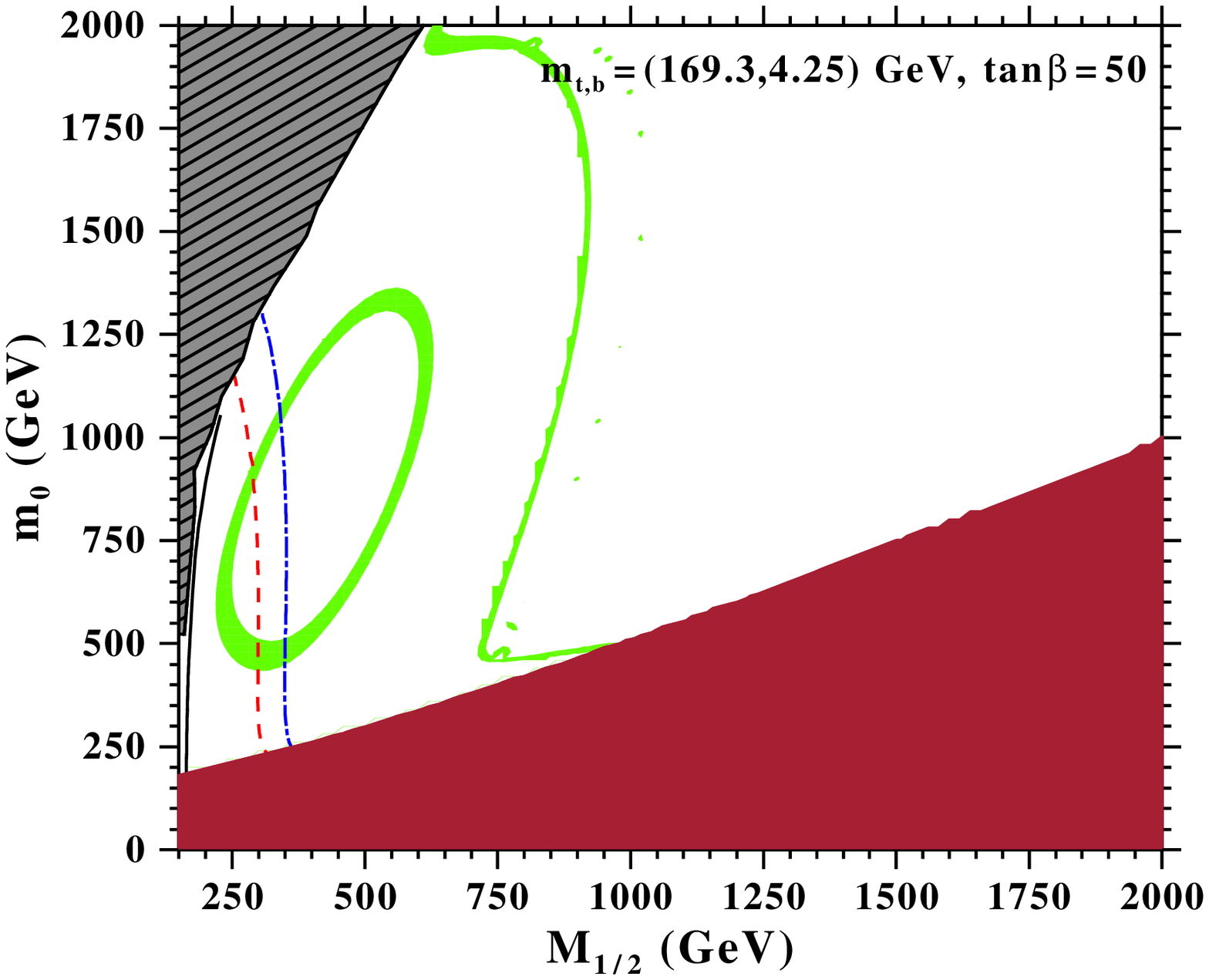}
\includegraphics[width=7.7cm]{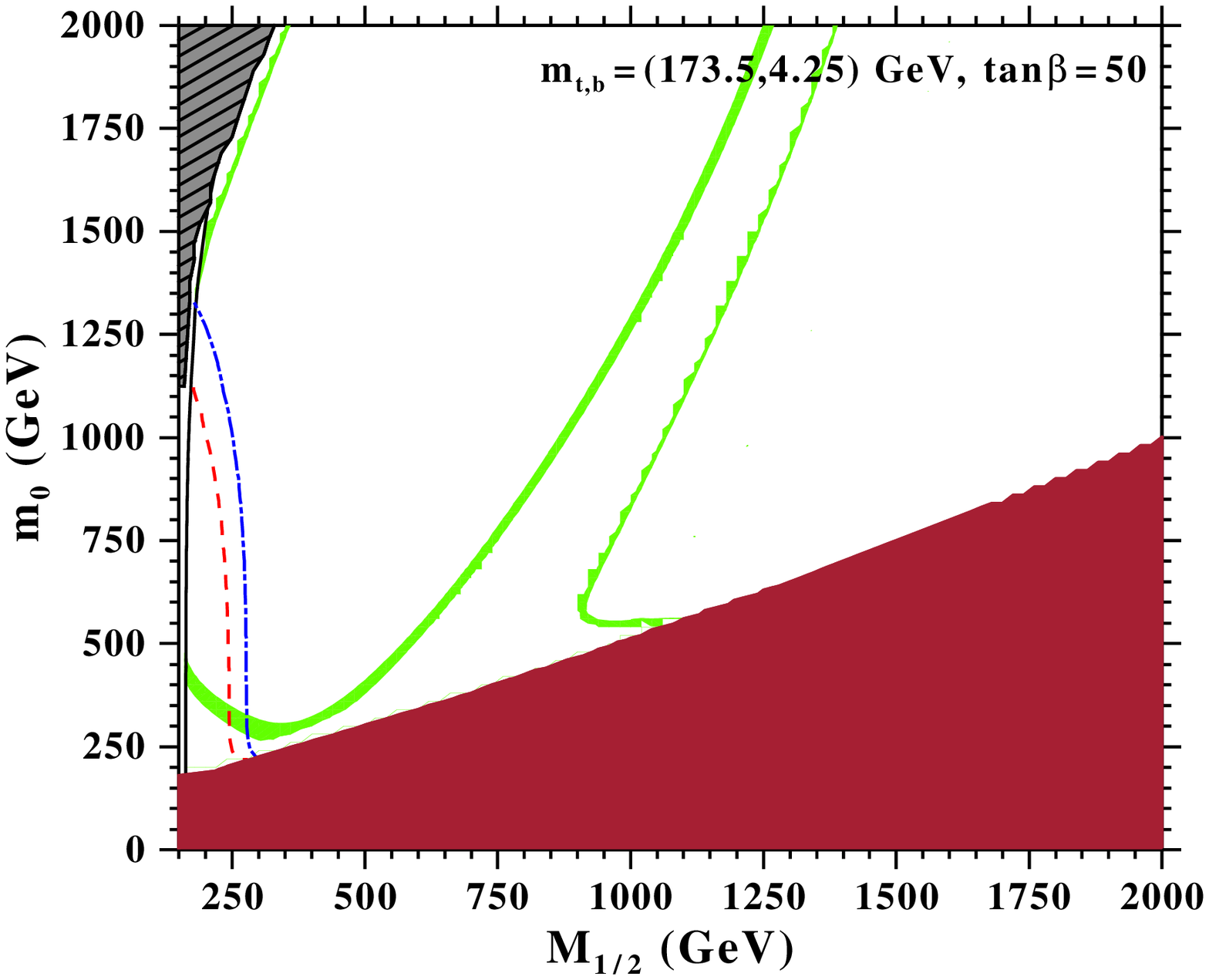}
\end{center}
\caption[]{The cosmologically  allowed  region (shaded light green)
 for $m_t=169.3$ GeV (left panel) and 
$m_t=173.5 $ GeV (right panel), when all  the phases are switched off, for $\tanb=50$. 
The magnitude of the common trilinear scalar coupling is taken $A_0=100$ GeV.
The remaining inputs are shown on the figure.
The solid black line on the left of the figure is the chargino mass 
bound $m_{\tilde{c}} > 105 $ GeV. 
The red dashed  (blue dashed-dotted) line  indicates the Higgs mass bound 
$114$ GeV ($115$ GeV).
The hatched area at the left-top designates the no-electroweak symmetry breaking region.  
At the bottom the  shaded region is excluded since there the stau is the LSP.}
\label{fig4}
\end{figure}

In the previous section we gave an account of the salient features of MFV models 
when CP is violated.  Our principal aim in this section is to explore regions of
 the parameter space in which the LSP neutralino relic density is within the stringent 
limits put by WMAP3, satisfying at the same time the EDMs and all available accelerator 
constraints. We focus our analysis on regions where the LSP neutralinos are paired 
annihilated through a Higgs resonance which is one of the prominent mechanisms in 
CP-conserving models, in order to obtain acceptable relic densities, although other
 regions of interest will be also explored. 
  Regions of the parameter space in which
 such a process is feasible, when  CP is conserved, have the shape of funnels lying on
 each side of the line on which $2 m_{LSP}/M_{A}$ is unity and this occurs for large 
values of $\tan \beta$. In the presence of CP violation, and especially for large values 
of the phases which we are interested in, the shape and the location of the funnels are 
different due to the impact of the phases on the corrections to bottom mass as has
 been discussed in the previous section. There are two cases which one should be 
interested in: 
%%%%%%%%%%%%%
\renewcommand{\labelenumi}{\Alph{enumi}.}
\renewcommand{\labelenumii}{\Alph{enumi}. \alph{enumii}}
\noindent 
%\begin{flist}{229}{MidnightBlue}
\begin{enumerate}
{\item{
  Cases where EDMs are naturally suppressed, without invoking any special mechanism, to
 lower the values of EDMs to acceptable levels having at the same time some of the phases 
large. These regions correspond to large $m_0, M_{1/2} > \;\cal{O} \mathrm{(few\; \TeV)}$.
 Certainly the EDMs can be suppressed if all phases are very tiny but this case is not
 physically interesting. }} 
{\item{
 Cases where EDMs are suppressed due to cancellations among the various contributions. 
This requires a tuning of the phases at low energies that have a large impact on EDMs. }}
\end{enumerate}
In general EDM constraints exclude a large portion in the $m_0, M_{1/2}$ plane 
allowing large values of  $m_0$ or $M_{1/2}$ so that the EDMs are naturally suppressed.
 Thus,  depending on inputs they cut off substantial part, or all, of the cosmologically 
allowed neutralino 
and stau coannihilation~\cite{coanni} tale and focus point region~\cite{fp} as well.  
At the same time they cut part of all of the cosmologically allowed region in which 
neutralinos annihilate through a Higgs resonance. 
These regions have the shape of funnels, whose location and form 
is sensitive to the top and bottom masses and may occupy regions allowed by the EDM 
constraints, if they happen to span large $m_0, M_{1/2}$ values. 
The sensitivity with the top mass is shown in Fig.~\ref{fig4}. 
On the left panel and for the inputs shown we display the cosmologically allowed 
region for $m_t=169.3  \GeV$, the lowest allowed by the experimental data for the top mass. 
On the right panel the same figure is shown with $m_t=173.5  \GeV$ which is the upper 
experimental limit. 
In both cases all phases are switched off but this behaviour holds in CP-violating 
cases as well.  
In the first case the cosmologically allowed regions, which follow 
the $M_{Higgs}/2 m_{LSP}=1$ curve, are not so peaked and occupy regions 
characterized by $M_{1/2} < 900 \GeV$. Increasing $m_t$, on the right panel, 
the location of the line $M_{Higgs}/2 m_{LSP}$,  which controls  the location 
and the shape of the rapid Higgs annihilation region, turns to the right 
dragging with it the cosmologically allowed region to higher $M_{1/2}$ becoming 
more pronounced and extended. 
This behaviour is due to the sensitivity of the Higgs mass spectrum with $m_t$.
The larger the top mass is the sharper the shape of the funnel, which extends 
towards large $m_0, M_{1/2}$ values, and larger the possibility of overlapping 
with EDM allowed regions. For the bottom mass the tendency is rather opposite
 and it is low values of $m_b$ that favour the formation of sharp cosmologically 
allowed funnels.

 %%%%
%%%%%%%%%%%%%%%%%%%%%%%%%%%%%%%%%%%%%%%%%%
%%%%%%%%%%%%%%%%%%%%%%%%%%%%%%%%%%%%%%%%%%%%%%%%%%%%%
%%%%%%%%%%%%%%%%%%%%%%%%%%%%%%%%%%%%%%%%%%%%%%%%
%%%%%%%%%%%%%%%%%%%%%%%%%%%%%%%%%%%%%%%%%%%%%
\begin{figure}[h!]
\begin{center}
\includegraphics[width=7.7cm]{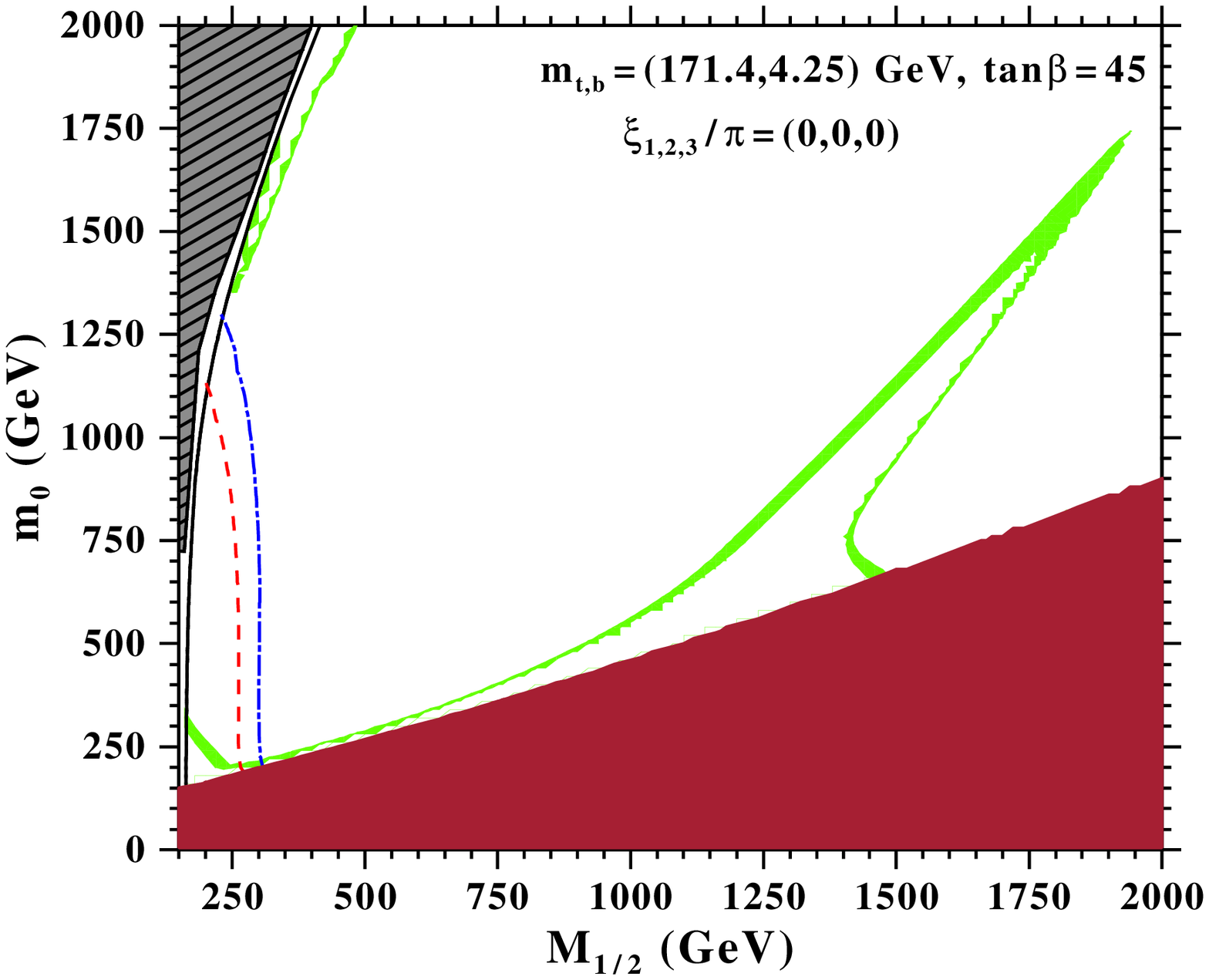}
\includegraphics[width=7.7cm]{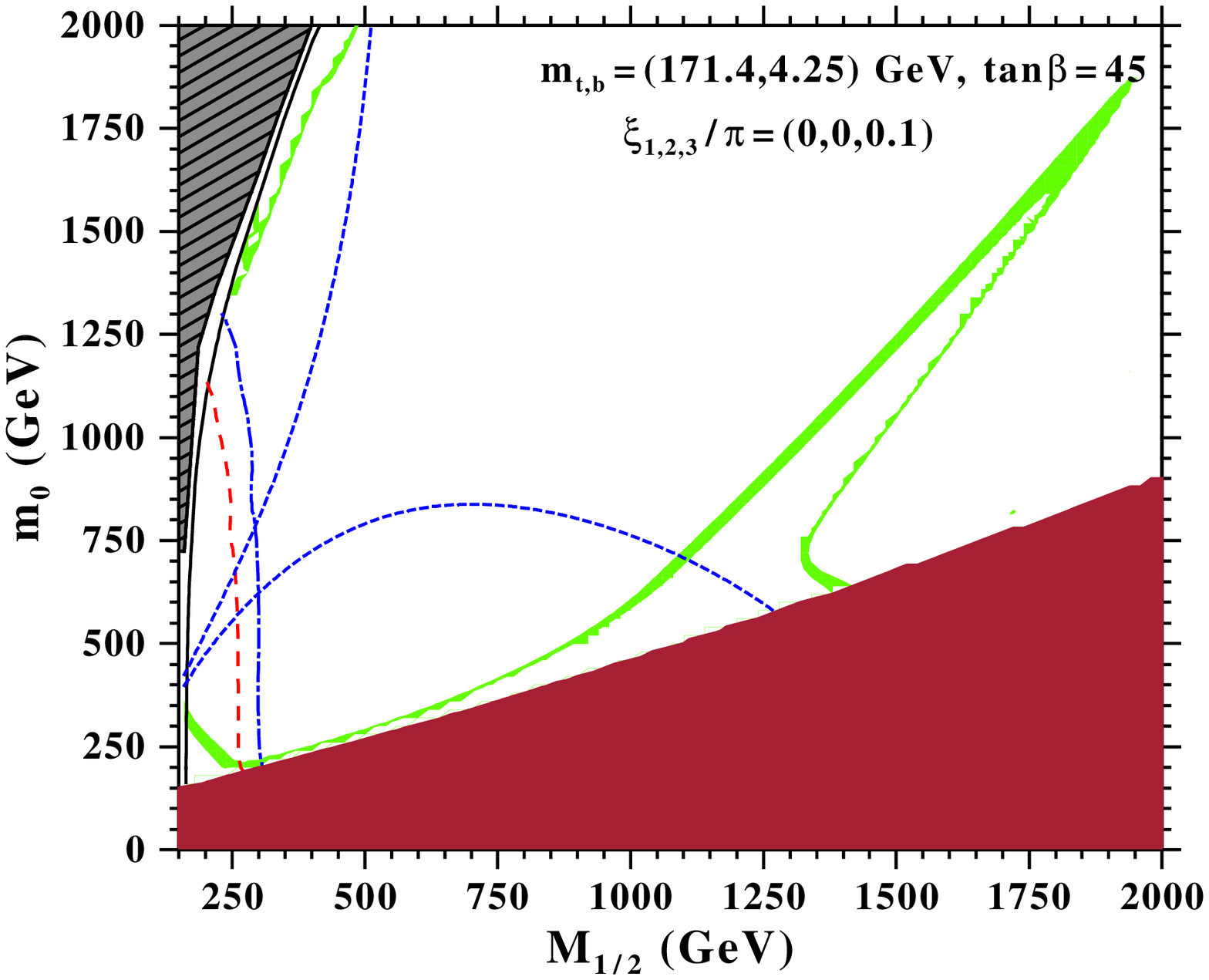}
\end{center}
\caption[]{The effect of the gluino phase alone for $\tanb=45$. On the right (left) panel
we take $\xi_3=0.1 \, \pi$ ($\xi_3=0$). The remaining  phases are zero and $A_0=100$ GeV.
On the right panel all displayed region is excluded by neutron and Hg EDM bounds. 
The allowed region by the electron EDM bound, $d_e=1.6 \times 10^{-26} e \cdot cm$, 
 lies  between the blue short-dashed lines. The rest of the curves and shaded 
regions  are as in Fig.~\ref{fig4}.}
\label{fig5}
\end{figure}
%%%%%%%%%%%%%%%

From the previous discussion it becomes evident that regions allowed by both EDM and
 DM constraints are easier to find  for the highest allowed values of the
 top mass and values of the phases that make the running bottom mass  minimum.
 In previous works \cite{Gomez:2005nr} the values of the top quark was taken as 
large as $178 \GeV$ in agreement with the experimental values quoted at that time. 
In view of the new experimental values of $m_t$, according to which $m_t$ 
is lowered by about $\sim 5  \GeV$, and due to the sensitivity on the top mass of the cosmologically 
funnels for LSP annihilation through a Higgs resonance , this picture may be distorted.
%%%%%%%%%%%%%%%%%%%%%%%%%%%%%%%%%%%%%%%%
It should be also remarked that in our analysis the situation is different from that 
encountered in \cite{Gomez:2005nr}  where Yukawa unification is enforced. In that 
case Yukawa unification entails to a different bottom mass value at each point of 
the $m_0-M_{1/2}$ plane and therefore due to the sensitivity on $m_b$ our 
findings cannot be directly compared to those of \cite{Gomez:2005nr}. 
%%%%%%%%%%%%%%%%%%%%%%%%%%%%%%%%%%%%%%%
%%%%%%%%%%%%%%%%%%%%%%%%%%%%%%%%%%%%%%

Before embarking on presenting our results we remark that for the 
calculation of the electric dipole moments and the Higgs masses we  
use the {\it{FeynHiggs-2.5.1}} code, \cite{FH}. However since $\vev{H_2}$ is not real, in order to 
implement the effect of the misalignment angle $\theta$,  
one has to replace the phase $\phi_{\mu}$ of $\mu$ by $\phi_{\mu}+\theta$.
For the electric dipole moments of the known species  
the correctness of this we have also checked numerically by comparing the outputs of our numerical routines against to those returned by  {\it{FeynHiggs}}.  
For the Higgs masses, in all cases studied, the Higgs mass spectrum obtained was close to that obtained by using the effective potential, with an accuracy $2-5 \%$. The latter is known to be less 
accurate, since the wave function renormalization effects are not counted for. Therefore in this work we use 
the Higgs masses as returned by  {\it{FeynHiggs}} and  
this comparison serves as a further  check of the correctness of our treatment. 
Concerning the experimental limits  put on Higgs masses, the ratio 
$\xi^2= {( g_{hZZ}/g_{hZZ}^{SM} )}^2$ of the light Higgs boson $h$ 
coupling to $Z$, $g_{hZZ}$,  to that of  the corresponding SM
coupling $g_{hZZ}^{SM}$ is considered for each case studied. 
This was found to be very close to unity, that is 
the light Higgs acts like a SM Higgs boson, and therefore the $LEP2$ 
limit $m_h > 114.5  \GeV$ applies.  This is expected in the constrained model 
studied in this work since we are within the decoupling region.
%%%%%%%%%%%%%%%%%%%%%%%%%%%%%%%%%%%

In case {\bf{A}} in order to locate regions compatible with all available data the 
funnel must be extended towards high $m_0, M_{1/2}$ entering regions in which EDMs 
are naturally suppressed. However, we have found that this cannot occur since 
EDM bounds require very high $m_0, M_{1/2}$ not overlapping with the cosmologically 
allowed funnel regions, unless, the CP-violating phases are very small. In order to 
conceive the picture, and for demonstrating the importance of a non-zero gluino 
phase, on the left panel of Fig.~\ref{fig5}  we display the cosmologically allowed 
regions, having the shape of funnels, when CP-violating phases are absent. 
The magnitude of the common trilinear scalar coupling is taken $A_0=100 \GeV$.
 The remaining inputs are shown on the figure. The hatched area on the left  
is not allowed due to the absence of EW symmetry breaking there. 
The solid black line on the left of the figure is the chargino mass 
bound $m_{\tilde{c}} > 105  \GeV$. The dashed line (in red) sets the Higgs mass 
boundary line $m_{Higgs} > 114  \GeV$, while to the right of it the dashed-dotted line 
(in blue) designates the bound on the Higgs mass reported by D0 \cite{:2007bxa}, 
$m_{Higgs} > 115  \GeV$. On the right panel for the same inputs we 
switch on  the gluino phase $\xi_3 = 0.1 \; \pi$. The funnel is slightly deformed
 with its top end approaching the point $m_0 = M_{1/2} = 2000  \GeV$. 
However all displayed region is excluded by neutron and Hg EDM bounds and it is not
 of physical relevance. The electron EDM, $d_e$, is also affected and allows the 
region confined between the short-dashed lines (in blue). 
This figure demonstrates in a clear way the impact of the 
gluino phase on $d_e$ by the two-loop RGE running of the phases, 
as discussed in the previous section. 
%%%%%%%%%%%%%%%%%

The other option opened, case {\bf{B}}, is to implement the cancellation mechanism 
according to which phases are chosen so that the various contributions to EDM 
cancel each other. This option does not require high values for $m_0, M_{1/2}$. 
Starting for such a point in the parameter space one can draw  trajectories, 
by merely rescaling the values of $m_0$ and $ M_{1/2}$, in which EDMs are in 
agreement with the experimental bounds put on them \cite{EDMcancelstring2}. 
These trajectories  may eventually overlap with the 
cosmologically allowed portions and thus delineate regions allowed by both 
EDM and cosmological constraints. 
In the previous section we discussed that this is rather difficult to be 
accomplished in the top-down approach due to the two-loop running of the
 phases and the interplay among these. 
Certainly  the phases can be tuned at the EW scale to cancel 
separate contributions in electron and neutron dipole moments. 
If such values are obtained at the EW scale, when one is 
subsequently trying to find extended regions  by rescaling as 
$m_0, M_{1/2} \rightarrow \lambda \; m_0, \lambda \; M_{1/2}$, 
the RGE evolution of these phases at the GUT scale yields values 
that depend on the parameter $\lambda$. Therefore their values at 
$M_{GUT}$ are fine tuned since they are different for different pairs of  
$m_0,M_{1/2}$ that are related by a rescaling factor. 
Therefore in the top-down approach  it is difficult to find extended 
regions in the $m_0, M_{1/2}$ plane 
by just rescaling the supersymmetry breaking parameters as prescribed before. 

In our approach, for given inputs for the supersymmetry breaking parameters, 
we prefer to implement this mechanism by rotating the input phases $\xi_{1,2,3}$ 
at the GUT scale until an acceptable point is found respecting the EDM limits 
on electron and neutron. Subsequently keeping fixed the values of the phases 
we vary $m_0, M_{1/2}$ to delineate regions compatible with all available data. 
Alternatively one can vary the 
phases around the values for which small EDMs for the electron and neutron were obtained,
keeping  the remaining inputs fixed, in order to locate regions in which large  phases 
are obtained satisfying all experimental bounds. The role of the phase $\xi_3$
 is vital in this approach. For fixed $\xi_2$ one varies $\xi_1$ until the 
electron EDM $d_e$ is small, by cancelling chargino against neutralino 
contributions. Subsequently $\xi_3$ is rotated until the neutron's EDM, $d_n$, is made small.

%%%%%%%%%%%%%%%%%%%%%%%%%%%%%%%%%%%
\begin{figure}
\begin{center}
\includegraphics[width=9.0cm]{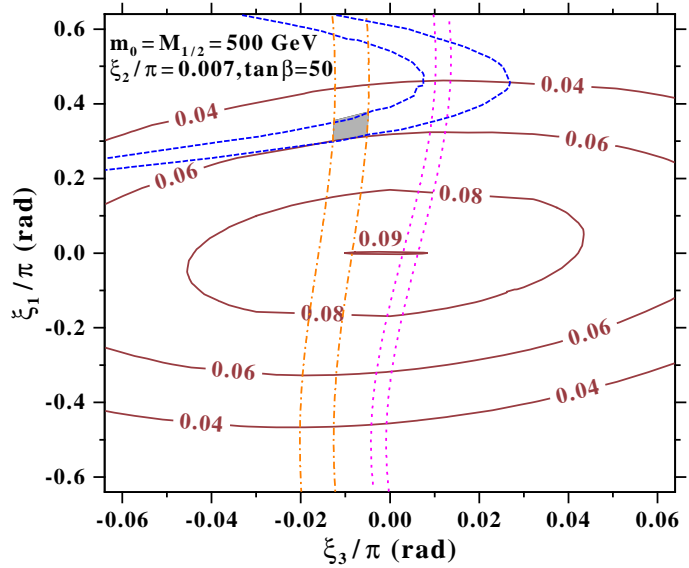}
\end{center}
\caption[]{ The EDMs and relic density contours 
for $m_0=M_{1/2}=500  \GeV$, $A_0=100$, $\tanb=50$,
and $\xi_2=0.007 \, \pi$ on the ($\xi_3,\xi_1$) plane. The remaining phases 
are zero at the GUT scale. The blue dashed (orange long dashed-dotted)
curves  designates the $d_e$ ($d_n$) acceptable region. 
The allowed region by $d_{Hg}$ is this between the magenta dotted  lines, which
does not overlap with the grey region allowed by electron and neutron EDMs.
The contours of constant neutralino relic density are shown as solid elliptical curves. 
The region favoured by WMAP3 data is located 
in  the centre of the figure  not overlapping with any of the the EDM  allowed domains. 
}
\label{fig_c1}
\end{figure}
%%%%%%%%%%%%%%%

This is the prescription followed in \cite{EDMcancelstring2} to make both
 $d_e$ and $d_n$ lie within their experimental limits. However as discussed 
in the previous section due to the two-loop RGE dependence of $d_e$ on $\xi_3$ 
the cancellation in $d_e$ may be lost and therefore $\xi_1$  need be re-rotated. 
It should be remarked that for large $\tan \beta$ the two-loop contributions are 
important  and cannot be ignored. In those cases therefore we apply this recipe 
by taking into account the appearance of two-loop contributions, as well as 
additional contributions from other sources, such as chromoelectric dipole 
and gluonic dimension-6 operators, which may be important. Therefore following this procedure 
one may be able to find points in the $\xi_{1,2,3}$ parameter space yielding small $d_e, d_n$. 
In this case however the Hg dipole moment  is not guaranteed to be within its experimental 
limits. Besides, since the cosmologically allowed regions depend on these phases, even 
if we start from a point on which the relic density is acceptable and implement the 
cancellation mechanism, we do not necessarily end up with phases that respect the 
cosmological bounds put on the relic densities. 
Such a situation is depicted in Fig.~\ref{fig_c1}. For the given inputs the
 phases $\xi_{1,2,3}$ are  rotated until the electron's and neutron's  EDMs 
are within their experimental limits. This point corresponds to $\xi_2 = 0.007 \; \pi$ 
and values of $\xi_{1,3}$ located at the centre of the small grey shaded region shown 
in the figure. This grey region is the overlap of the two stripes, one between the 
dashed lines, which designates the $d_e$ acceptable region, and the other between 
the dotted-dashed lines which designates the region allowed by the neutron EDM bounds. 
Starting from this particular $\xi_{1,2,3}$ point, and keeping $\xi_2$ fixed, we have 
plotted in the $\xi_{1,3}$ plane the allowed domains by EDM and relic densities. 
The allowed region by the Hg dipole moment is confined  between the dotted  lines (in magenta) 
which, as one can see, does not overlap with the grey region allowed by electron and neutron EDMs. 
In this figure it is also seen how the phases are fine tuned, especially $\xi_3$,  
to achieve acceptable EDMs for the electron and neutron in the sense that only in a 
small region both $d_e$ and $ d_n$ can be within their experimental bounds. 
On the same figure the contours of constant neutralino relic density are shown as 
solid elliptical curves. In the example shown,  
one observes that the location of the region favoured by WMAP3 data 
occupies only a small portion at the centre of the figure at which $\xi_{1,2} \simeq 0.0$ 
not overlapping with the EDM  allowed domains. This figure represents  a rather typical 
example of the difficulty one encounters to reconcile both EDMs and cosmological bounds.

\begin{figure}[t]
\begin{center}
\includegraphics[width=7.7cm]{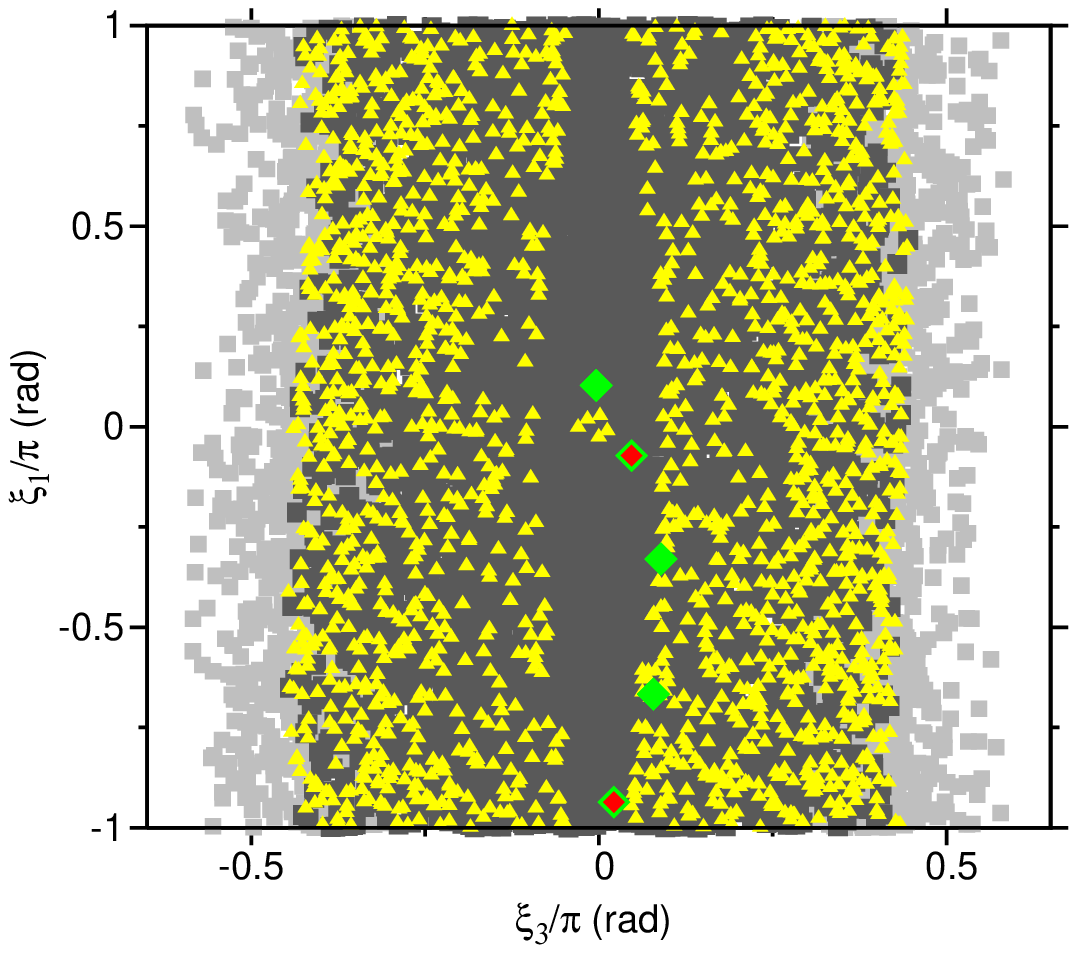}
\includegraphics[width=7.7cm]{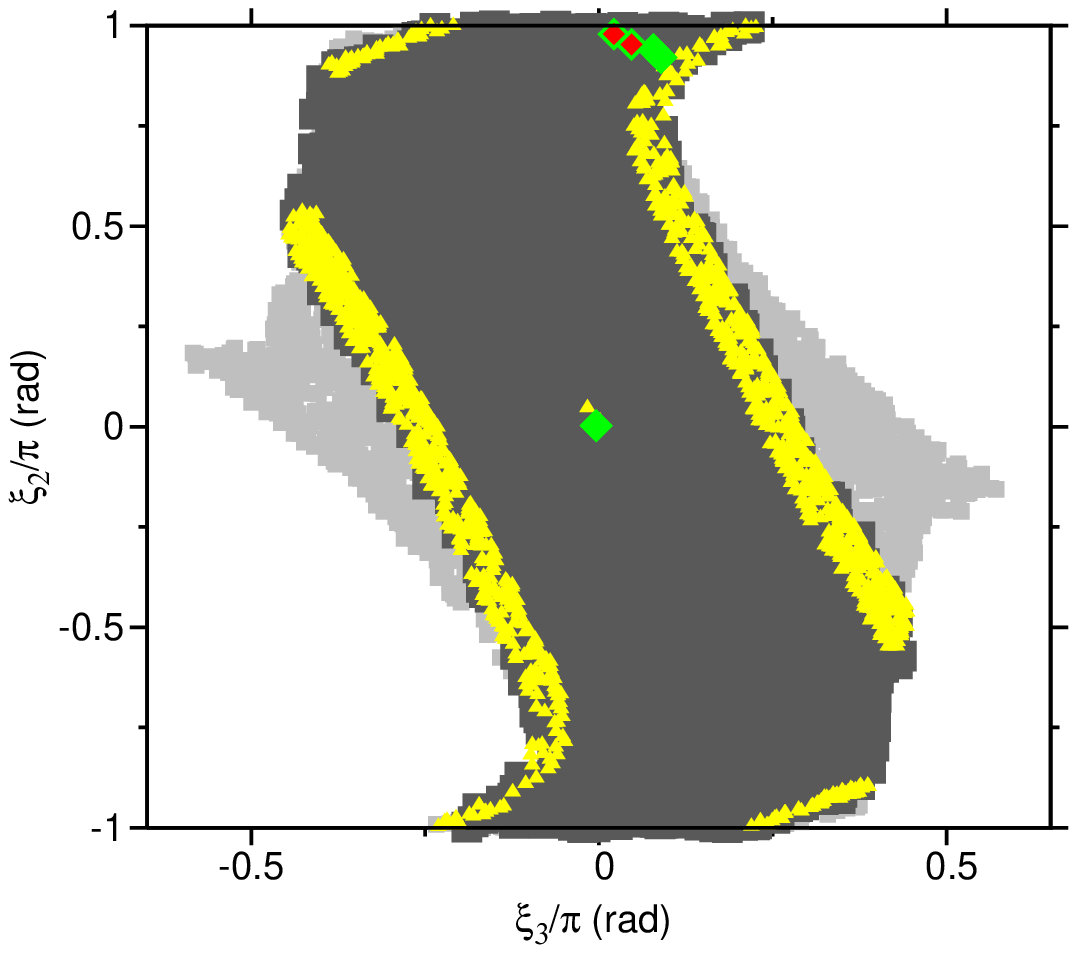}
\end{center}
\caption[]{Scatter plots in $\xi_3, \xi_1$ (left panel) and $\xi_3,\xi_2$ 
(right panel) plane, based on a random sample of the gaugino phases $\xi_{1,2,3}$ for fixed 
$m_0= 500  \GeV$, $M_{1/2}=480  \GeV$, $A_0=100 \GeV$ and $\tan \beta=50$. 
The remaining phases  at the unification scale are taken zero. 
The light grey squares represent  points that satisfy  the mass bounds for the 
light Higgs boson
and other SUSY particles. The dark grey region is formed 
from  points that in addition they satisfy the upper WMAP3 bound  for the relic 
density $\Omega_{\tilde \chi} h_0^2 < 0.117$. The yellow triangles represent 
their subset falling within the WMAP3 region, 
$0.089 < \Omega_{\tilde \chi} h_0^2 < 0.117 $.
The one-colour (green) diamonds satisfy  the electron and neutron EDM bounds,
while those filled with different colour (red), satisfy, in addition, the Hg 
EDM bounds. Depending on which region they lie on, they may observe the bounds put by WMAP3. 
}
\label{fig7}
\end{figure}

In Fig.~\ref{fig7} we have taken moderate values  $m_0=500  \GeV$, $M_{1/2}=480   \GeV$, 
$A_0=100  \GeV$. and  large $\tan \beta=50$. The masses of the top and bottom 
are $m_{t, b}=171.4, 4.25   \GeV$. The figure is constructed from one million 
random points in the $\xi_{1,2,3}$ parameter space. All other phases are assumed 
zero at the unification scale. Initially a particular point, 
$\xi_{1,2,3}= -0.072 \pi, 0.953 \pi, 0.047 \pi$ is found  by tuning the phases, in 
the way described earlier, satisfying all EDM bounds including in this case the EDM 
constraints from Hg atoms as well. The random sample is chosen to include this 
particular point. In Fig.~\ref{fig7} we classify these points in the planes $\xi_{1,3}$ 
(left panel) and  $\xi_{2,3}$ (right panel) according on what bounds each 
point satisfies. Light grey squares, forming the dense light grey region (region-1), 
includes points that satisfy only the Higgs and chargino mass bounds. The subset of 
these, shown as dark grey region (region-2), includes points that in addition they 
satisfy the upper observational limit for the relic 
density, i.e. $\Omega_{\tilde \chi} h_0^2 < 0.117 \;$ 
but they do not necessarily fall within the region dictated by the WMAP3 region. 
Their subset, marked as triangles (region-3), are within the $2 \sigma$ 
WMAP3 limits for CDM,  $0.089 < \Omega_{\tilde \chi} h_0^2 < 0.117 \;$. 
For the rest $\Omega_{\tilde \chi} h_0^2 < 0.089 \;$
and therefore they are of relevance if additional components, except the 
LSP neutralino, contribute to the total DM density.

%%%%%%%%%%%%%%
%%%%
%%%%%%%%%%%%%%
\begin{figure}[h!]
\begin{center}
\includegraphics[width=7.7cm]{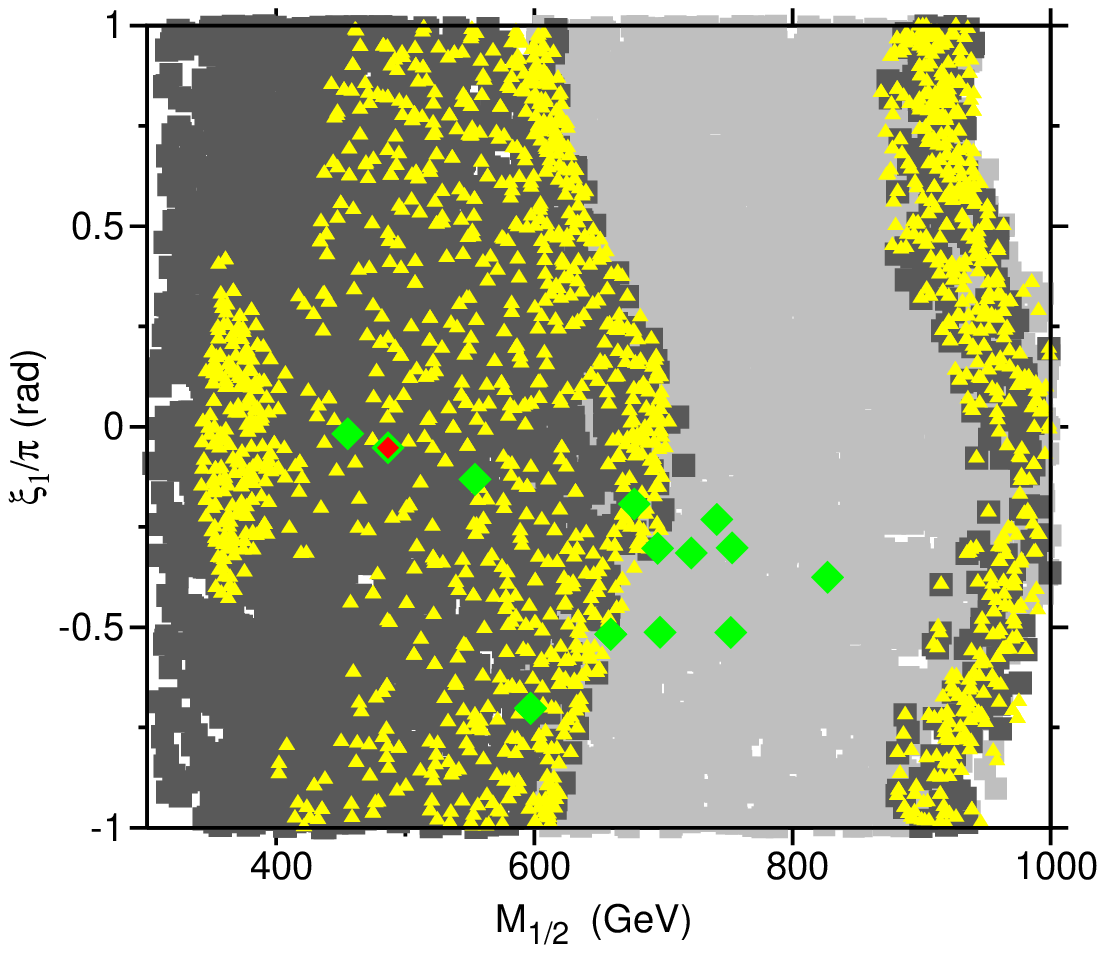}
\includegraphics[width=7.7cm]{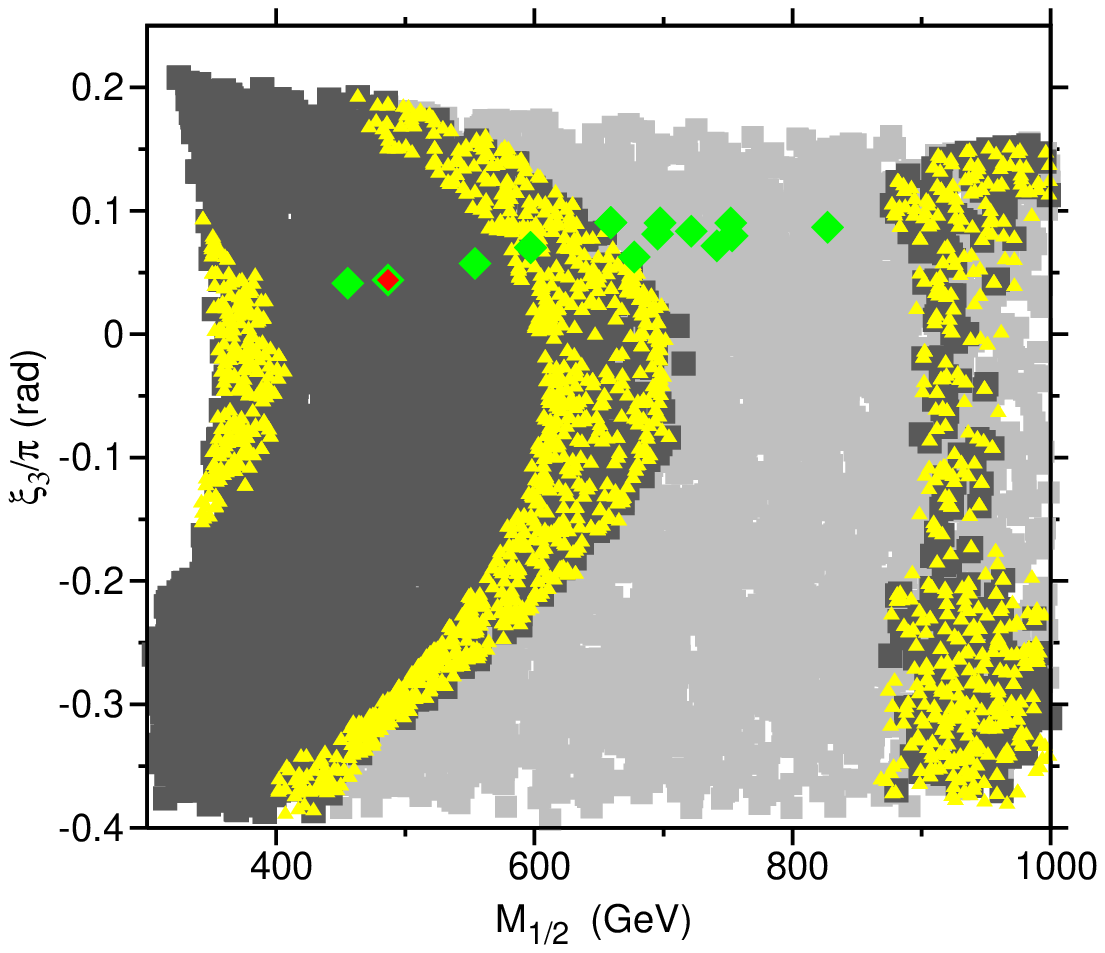}
\end{center}
\caption[]{Scatter plots in $\xi_1,M_{1/2}$ (left panel) and 
$\xi_3,M_{1/2}$ (right panel) plane, for a random sample with $\xi_2$ fixed and 
random values for $\xi_{1,3}$ and $M_{1/2}$. 
The notation is  as in Fig.~ \ref{fig7}. 
}
\label{fig8}
\end{figure}
%%%%

The points shown as coloured-filled diamonds satisfy the electron and neutron EDM bounds.
 Of those only their subset filled with different colour (red) satisfy, in addition, the Hg 
EDM bounds. If any of these points falls on regions-1,2 or 3, defined before, then in 
addition it satisfies the corresponding bounds designating this particular region. 
One observes that irrespectively of the cosmological, and other accelerator data, 
the EDM bounds by themselves are hard to satisfy for moderate $m_0, M_{1/2}$. 
Despite the fact that a particular point was found which respects all three EDM bounds, 
the random sample of one million points in the $\xi_{1,2,3}$ space leaves only a few 
that observe the EDM limits for electron and neutron and even fewer that satisfy all 
three EDM constraints. This demonstrates that the values of the phases must be fine 
tuned to comply  with experimental data of electric dipole moments. In the examples 
shown none of the diamond points satisfy the WMAP3 cosmological constraints. In fact 
for these points the predicted neutralino relic density is below the limits put by 
WMAP3 and therefore additional DM candidates must exist to fill the deficit. 

In Fig.~\ref{fig8}, with the same starting values for the phases $\xi_{1,2,3}$ and the 
parameters
$m_0$, $M_{1/2}$, $A_0$, $\tan \beta$, for which agreement with all EDM bounds are 
obeyed, we keep $\xi_2$ fixed and generate a random sample of one million points for 
$\xi_{1,3}$ and $M_{1/2}$. The sample includes the 
starting values for $\xi_{1,3}, M_{1/2}$. 
In $\xi_1,M_{1/2}$ (left panel) and $\xi_3,M_{1/2}$ (right panel) planes, the points 
satisfying the particular criteria as described in Fig.~\ref{fig7} are displayed. The
 notation is  as in Fig.~\ref{fig7}. The points satisfying the electron and neutron 
EDM bounds are few, and only one satisfies all three EDM bounds. In the case considered a 
few of these points overlap with the region-3 (yellow triangles) and therefore for these 
all available data are obeyed, with the exception of the Hg EDM bound. The limited number 
of points satisfying the EDM bounds, in this case too, it indicates that the phases must 
be fine tuned to agree with experimental data.

\begin{figure}[t]
\begin{center}
\includegraphics[width=7.7cm]{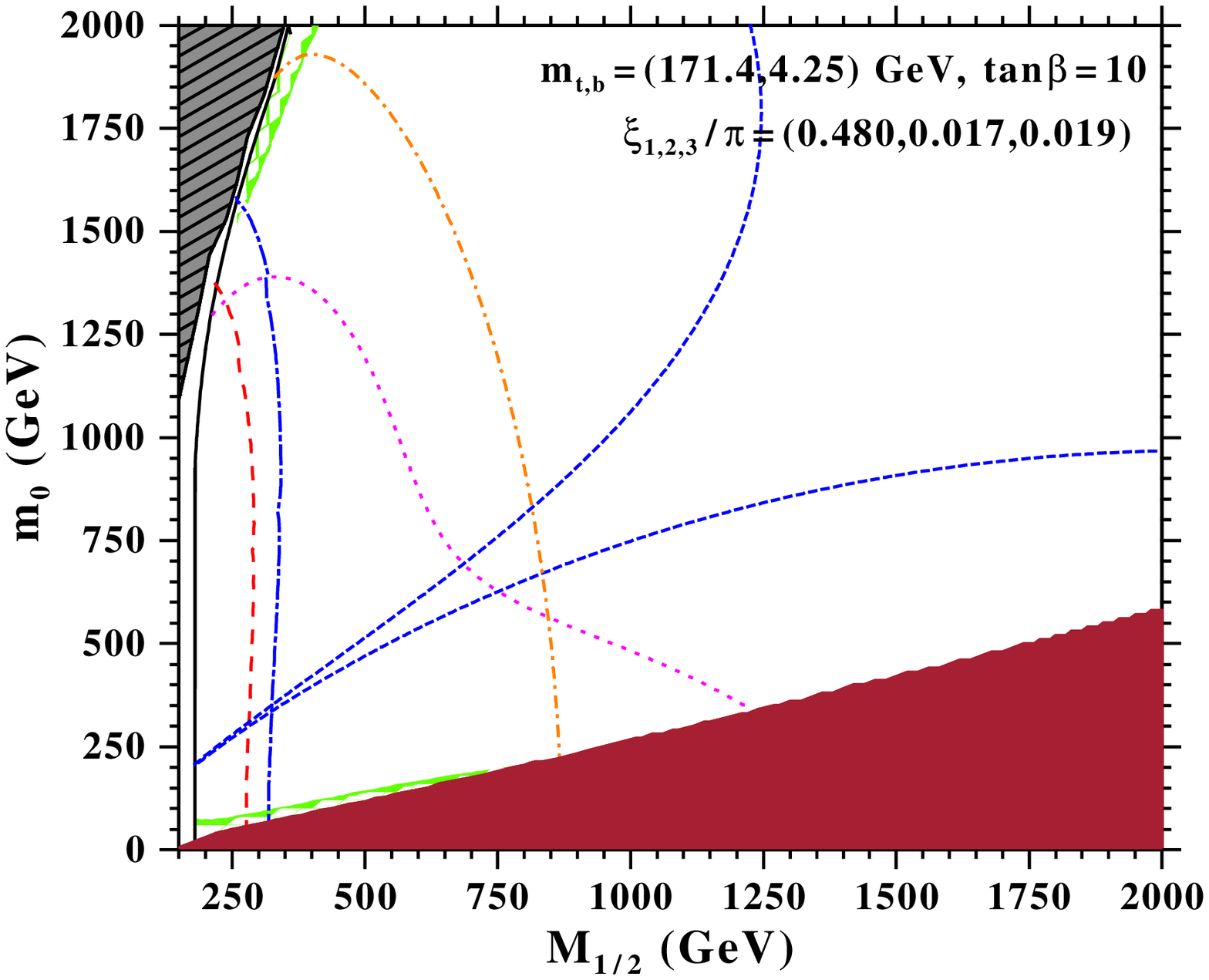}
\includegraphics[width=7.7cm]{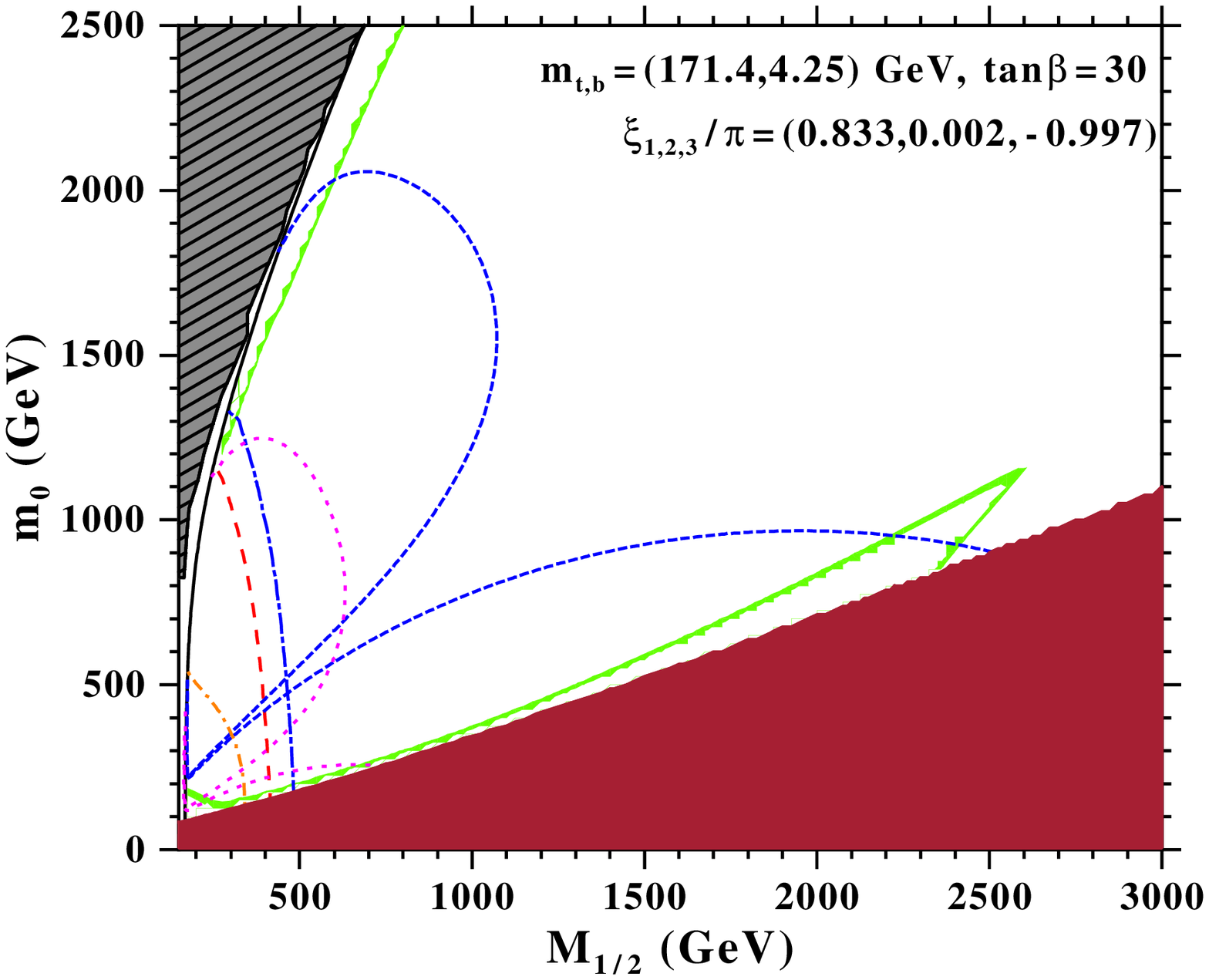}
\end{center}
\caption[]{
The $m_0,M_{1/2}$ parameter space for non zero gaugino phases, with values  
shown on the figures.
$A_0=100 \GeV$  and the phase of $\mu$ is taken vanishing. The dashed lines 
(in blue) delineate the boundaries of $d_e$, the dotted  lines 
(in magenta) those of $d_{Hg}$ and the dotted-dashed lines (in orange) 
those of $d_n$. If only one boundary line is shown it means the other 
one lies outside the displayed $m_0, M_{1/2}$ range. 
The allowed by EDM constraints regions are between the boundaries in each case. 
The rest of curves and shaded regions are as in Fig.~\ref{fig5}.}
\label{fig9}
\end{figure}

\begin{figure}[h!]
\begin{center}
\includegraphics[width=7.7cm]{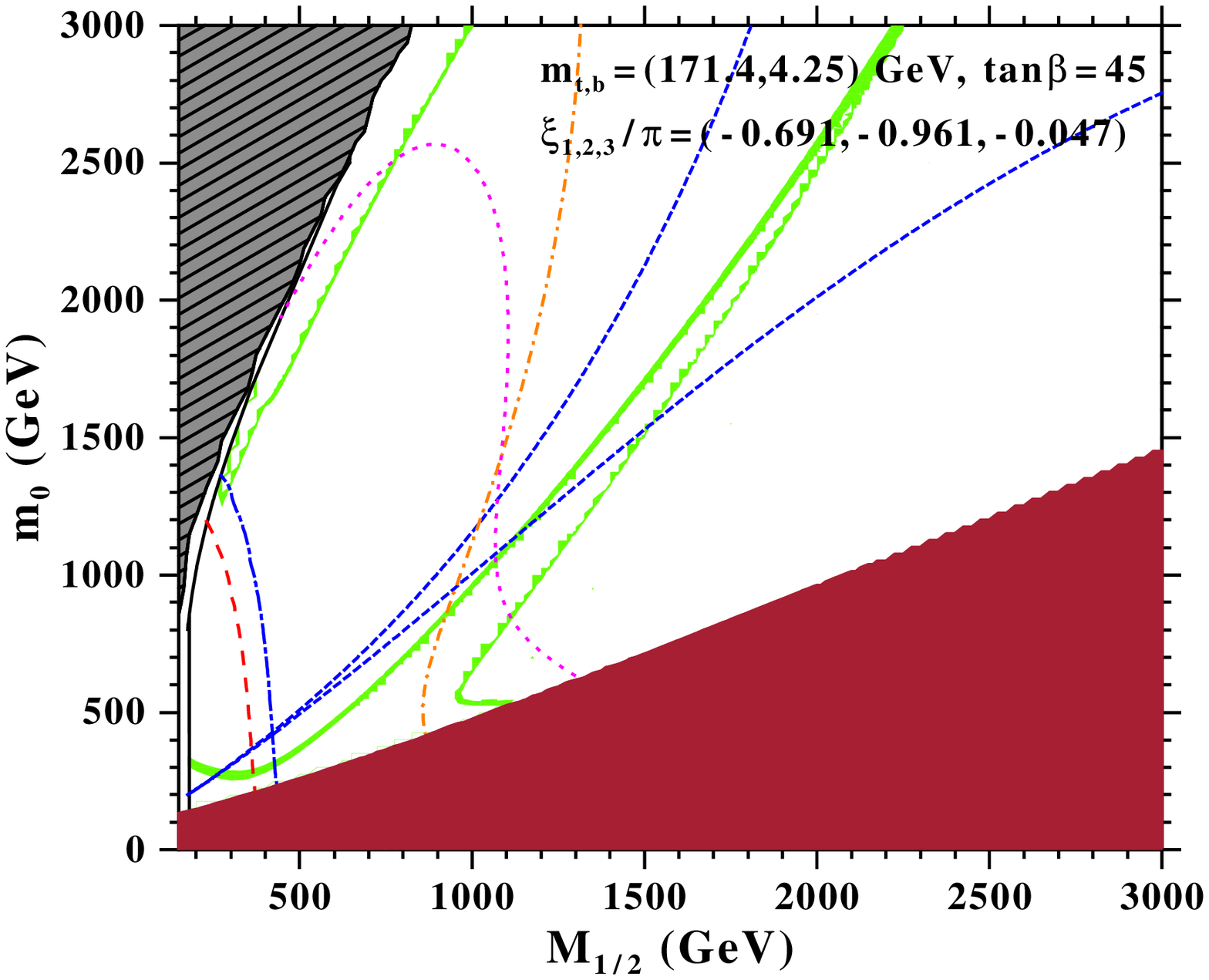}
\includegraphics[width=7.7cm]{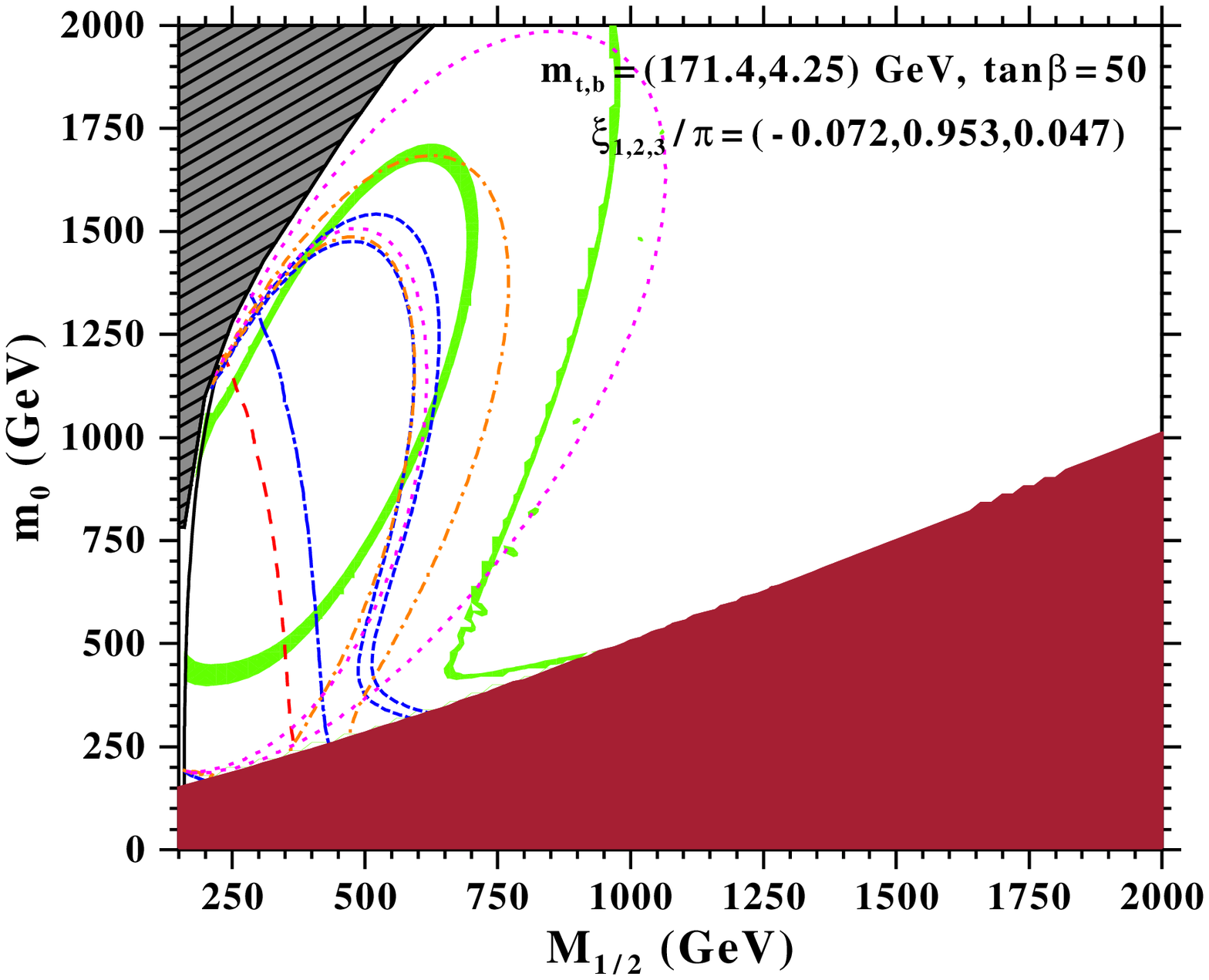}
\end{center}
\caption[]{ As in Fig.~\ref{fig9} for different values of $\tan \beta$ and the gaugino phases.
}
\label{fig10}
\end{figure}

The Hg EDM bound, in general, poses a severe obstacle in  obtaining agreement with 
cosmological data. However by tuning  appropriately the gaugino phases there are 
cases where agreement with all EDMs is obtained at a particular 
point $m_{0},M_{1/2}, A_0$ of the parameter space.  
Then by varying $m_0,M_{1/2}$ there is a chance that one succeeds in obtaining 
cosmologically and EDM allowed regions which overlap. The chance of obtaining this is 
increased if one considers funnel regions of  rapid neutralino annihilation via a
 Higgs resonance,  which occupy extended regions in the parameter space. 
In figures \ref{fig9} and \ref{fig10} we display cases for low and large $\tan \beta$. 
The values for the phases $\xi_{1,2,3}$, in each figure, are fine tuned for 
specific $m_0,M_{1/2}$, around $600 \GeV$,  in order to obtain agreement with all three EDMs, 
 $d_e$, $d_n$, $d_{Hg}$, for the electron, neutron and Hg respectively. 
We have taken $A_0=100   \GeV$  and the phase of $\mu$ is taken vanishing. All other 
inputs are shown on the figures. The dashed lines (in blue) delineate the boundaries 
of $d_e$, the dotted  lines (in magenta) those of $d_{Hg}$ and the dotted-dashed 
lines (in orange) those of $d_n$. If only one boundary line is shown it simply  means 
the other one lies outside the displayed $m_0, M_{1/2}$ range. 
The allowed by EDM constraints regions are between the boundaries in each case. In 
all 
cases displayed there are regions where all electric dipole bounds are simultaneously 
satisfied. Also shown are the Higgs line $114.0 \GeV$, long-dashed (red), and 
the line $115.0  \GeV$,  long dashed-dotted (in blue), lying on the left   
almost vertical to the $M_{1/2}$ axis. Their upper ends touch  the 
no-electroweak symmetry breaking region, designated as a hatched area.
 At the bottom the triangle-shaped shaded region is excluded since there the stau \;
is lighter than any of the neutralinos. The $2-\sigma$  cosmologically allowed WMAP3 regions are shown as shaded contours 
(in light green).

%%%%%%%%%%%%%%
%%%%
%%%%%%%%%%%%%%%
%%%%%%%%%%%%%%%

On the left panel of Fig.~\ref{fig9} the value of $\tan \beta=10$ is small and 
the cosmologically allowed regions are not so extended. Although for the specific 
values of the gaugino phases there are large portions of the parameter space
 compatible with all EDMs, the overlap of these regions with the regions allowed 
by the WMAP3 data extend to large values of $m_0 > 5.0 \TeV$, not shown in the figure, along the focus point region \cite{fp}.  
On the right panel of the same figure the value of $\tan \beta=30$ is considerably 
larger and the cosmologically allowed domains have a larger 
overlap with the regions allowed by EDMs. In this case there is a region in which 
all data are satisfied for $m_{0} \sim 1 \TeV$
and $M_{1/2} > 2.2 \TeV$ due to the fact that the allowed by $d_e$ domain is enlarged, 
allowing for smaller $M_{1/2}$ values, which includes part of a funnel that starts being formed. 
This is located on the right of the figure, just above the shaded region which is excluded since there the stau is the LSP. 
Due to the heaviness of $M_{1/2}$ this is outside the reach of LHC. 
As in the left panel case there is also a focus point region, acceptable by all data, starting now from smaller values of $m_0 > 2.1 \TeV$, which tracks the border of the no-electroweak breaking region. Part of it includes points with $M_{1/2} < 800 \GeV$ being therefore accessible to LHC. 

Notice that the values of the phases for the right panel
are different from those of the case displayed on the left. Due to the fine
 tuning of the phases chosen  increasing the value of $\tan \beta $ from 10 to 
30 in the left panel we do not get a picture resembling the one we display on
 the right panel. 

In Fig.~\ref{fig10}, for different sets of the gaugino phases and large values
 of $\tan \beta$, we display regions that are allowed by all data. 
Again the phases have been fine-tuned and are different for each 
case shown in the two panels. 
On the left panel the value of $\tan \beta$ is large, $\tan \beta =45$,  and the cosmologically 
allowed domains are funnel-shaped,  extended diagonally towards high 
$m_0$ and $M_{1/2}$ values, $m_0 \simeq 3.0 \TeV, M_{1/2} \simeq 2.2\TeV$. At the same time the acceptable EDM domains are also extended having a large overlap with the cosmologically allowed region. 
In this particular case there is a large portion of the parameter space for 
$m_0, M_{1/2} > 1.3 \TeV$ in which all experimental data are satisfied.
 A small part of this region is within the reach of LHC. 
The conclusion from this is that by tuning the gaugino phases there can be found  
extended regions in the parameter space in which rapid neutralino annihilation 
through a Higgs resonance can coexist with regions allowed by the stringent 
constraints imposed by the electric dipole moments. 
On the right panel the value of $\tan \beta=50$ is larger and it also allows for the 
formation of rather extended cosmologically allowed regions which are almost
 funnel-shaped, but no so peaked as the case considered previously. The boundaries of EDMs in the case shown lie within the range 
$m_{0}< 2.0 \TeV,M_{1/2} < 1.0 \TeV$ and one observes that EDM and cosmologically allowed domains 
again overlap. The overlapping regions are not so extended however, as in the case 
considered previously. In the case shown it is only a small region centered around 
the point $m_0, M_{1/2}= 900, 600  \GeV$. 
One observes that larger values of $\tan \beta$, unlike the CP-conserving case, 
do not always correspond to funnels spanning the highest $m_{0,1/2}$ regions. 
This is due to the sensitivity of the bottom mass, and hence the funnel regions, 
with the CP-violating phases.  
As an effect cosmologically allowed funnel-shaped regions can appear for smaller 
values of 
$\tan \beta$, as compared with the CP-conserving case, and can coexist with EDM 
allowed domains.  

%%%%%%%%%%%%%%%
\begin{figure}[t]
\begin{center}
\includegraphics[width=7.7cm]{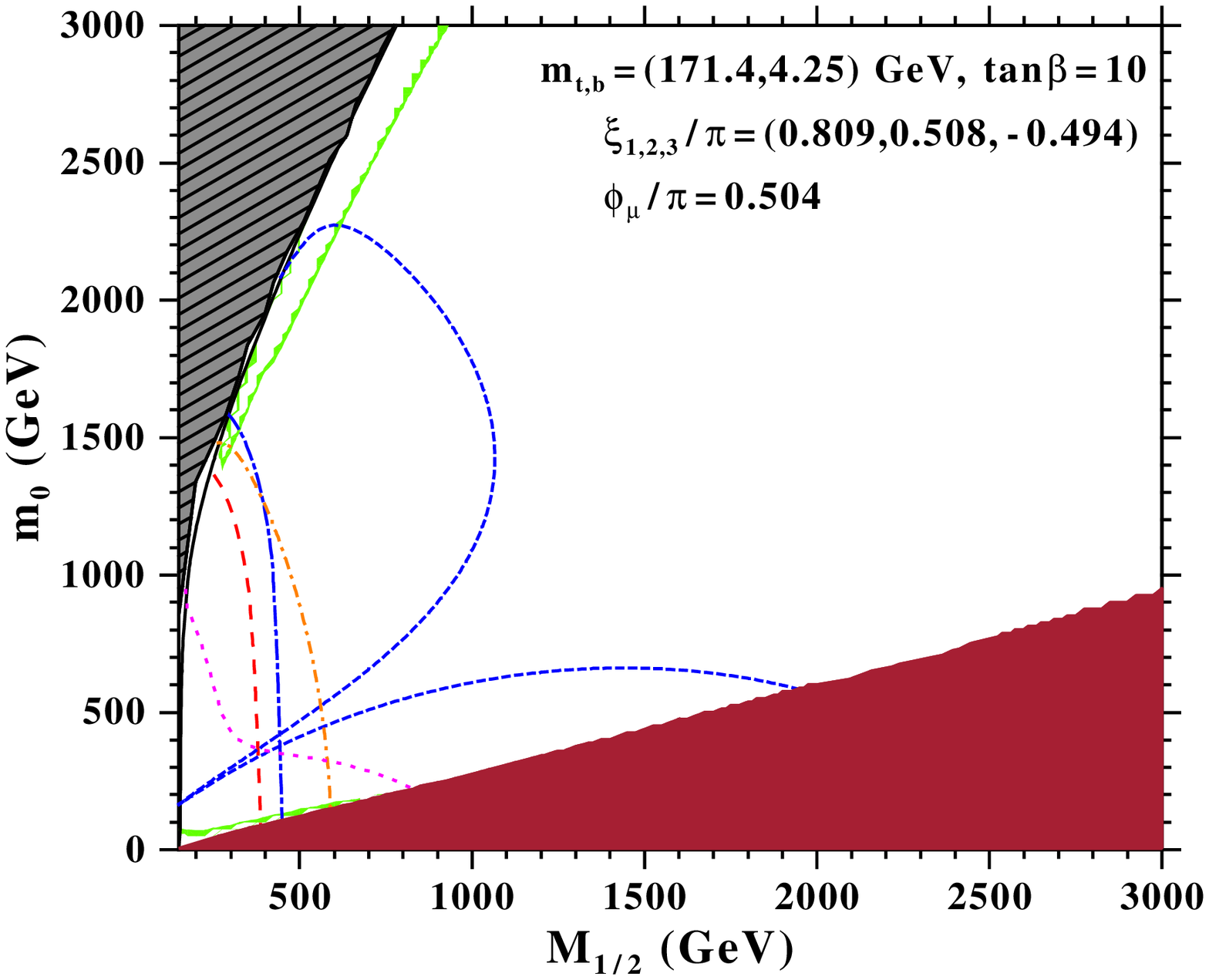}
\includegraphics[width=7.7cm]{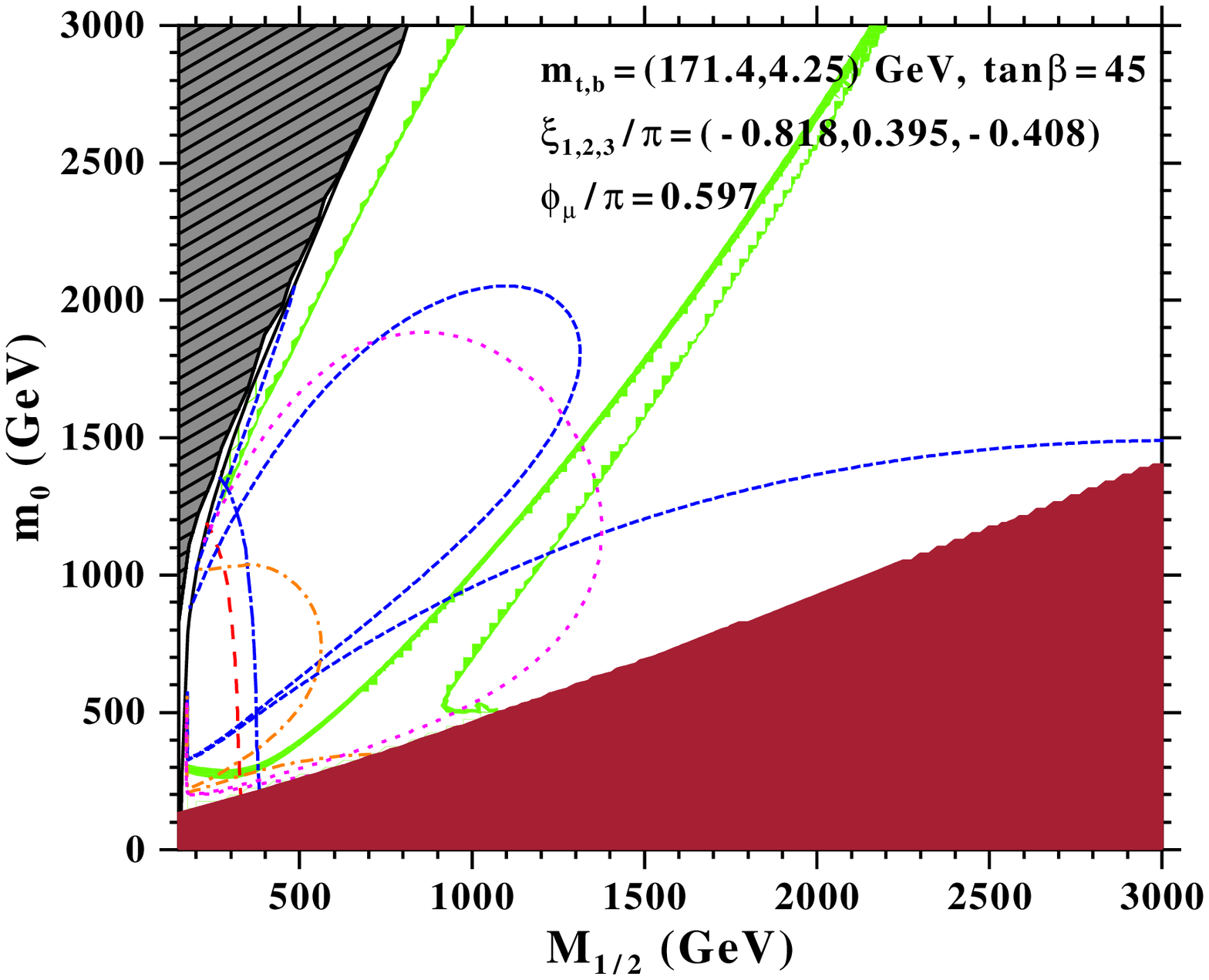}
\end{center}
\caption[]{As in Fig.~\ref{fig9}, with a non-vanishing phase of $\mu$
and values of the gaugino phases and $\tan \beta$ shown in the panels.
}
\label{fig11}
\end{figure}
%%%%%%%%%%%%%%%

In the previous analysis the $\mu$ phase has been put to zero but one can 
also seek for cases where 
large values of this phase are allowed. This is forbidden in cases where the 
gaugino phases are switched off. In mSUGRA models, in which the common phase 
of the trilinear coupling and the $\mu$ phase are the only allowed 
supersymmetric CP-violating sources, it is known that the phase of $\mu$ 
is tightly constrained by the EDM data, especially by this of the electron. 
However when one allows for the 
presence of different gaugino phases the situation is altered and this 
restriction is relaxed. In Fig.~\ref{fig11} we display a case where $\phi_\mu$ 
is non-vanishing and large. On the left panel a case with $\tan \beta =10$ 
is shown. On the right panel 
$\tan \beta =45$. The value of $A_0$ is $A_0=100\GeV$.
In both cases the phase of $\mu$ is non-vanishing and there 
are regions compatible with EDMs, cosmological data, as well as all other 
accelerator data. 
On the left panel only the focus point region is compatible with all data, while on the right panel  both focus point region and the cosmologically allowed funnels are allowed. 
Both regions include points accessible to LHC, which are larger as compared to the cases considered previously.

Note that the values of $arg(\mu M_{1,2})$ at $M_{GUT}$ 
are sizable, $ {\cal{O}} ( {0.1} \; \pi)$. Due to absence of renormalization 
of the $\mu$ phase with the energy scale and the small renormalization of
 the gaugino phases, which at one-loop are independent of the SUSY inputs at $M_{GUT}$, 
 these quantities retain almost the same values at the EW scale $M_Z$,
 for every point in the $m_0, M_{1/2}$ plane. The difference of their values at
 $M_Z$ from those at $M_{GUT}$ is at per mille level which is very small and 
important only  for EDMs. 
As stated in the introduction 
these combinations of phases set the measure of sufficient Higgsino and gaugino 
driven Baryogenesis at the EW phase transition. On these grounds and 
in conjunction with the fact that the magnitude of $\mu$ is comparable to that 
of $M_{1,2}$ in a large region of the allowed parameter space, displayed in 
this figure, these regions may be also compatible with these 
Baryogenesis mechanisms \cite{charEWB,neutrEWB,resonant}.  

%%%%%%%%%%%%%%%
\begin{figure}[t]
\begin{center}
\includegraphics[width=7.7cm]{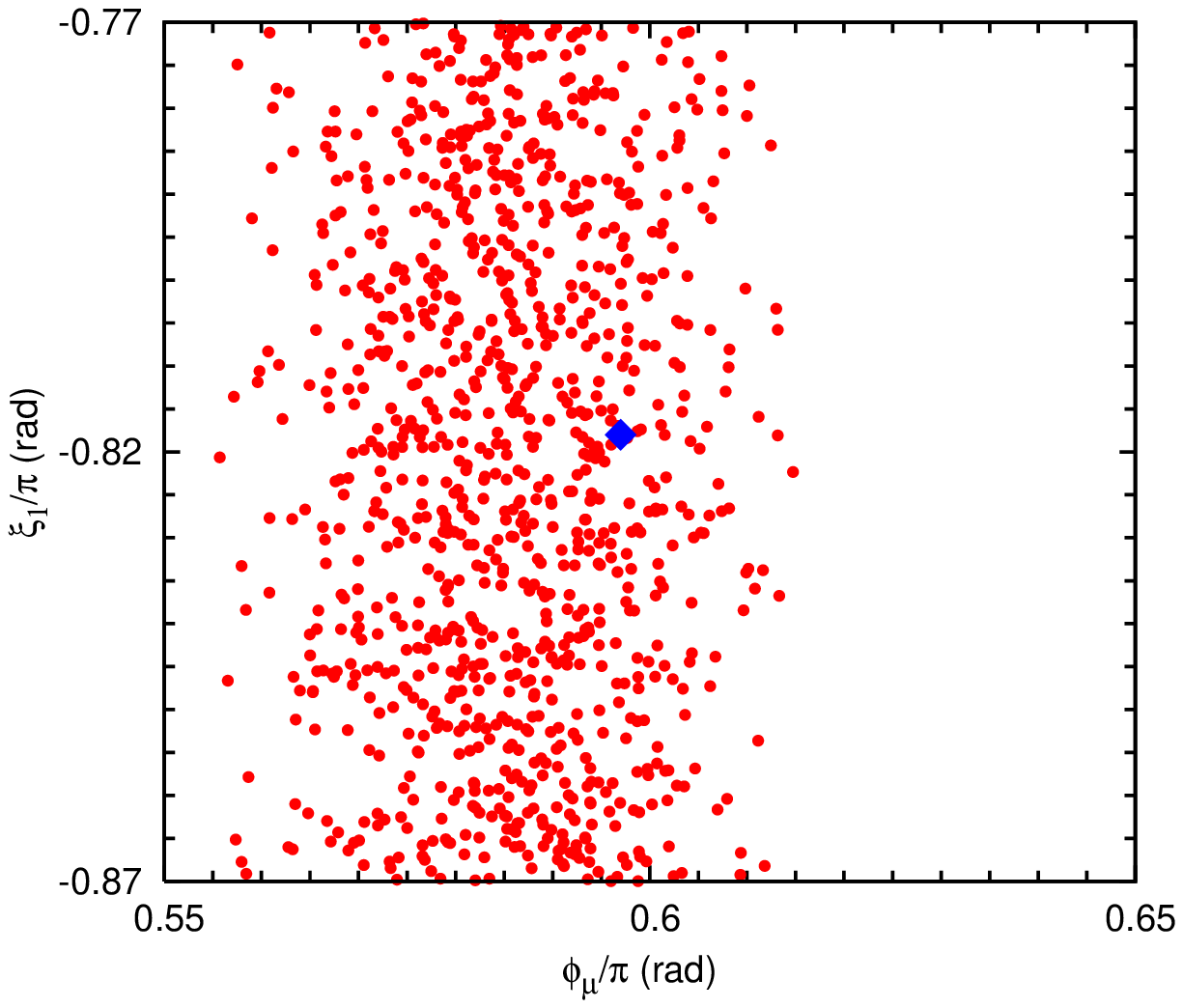}
\includegraphics[width=7.7cm]{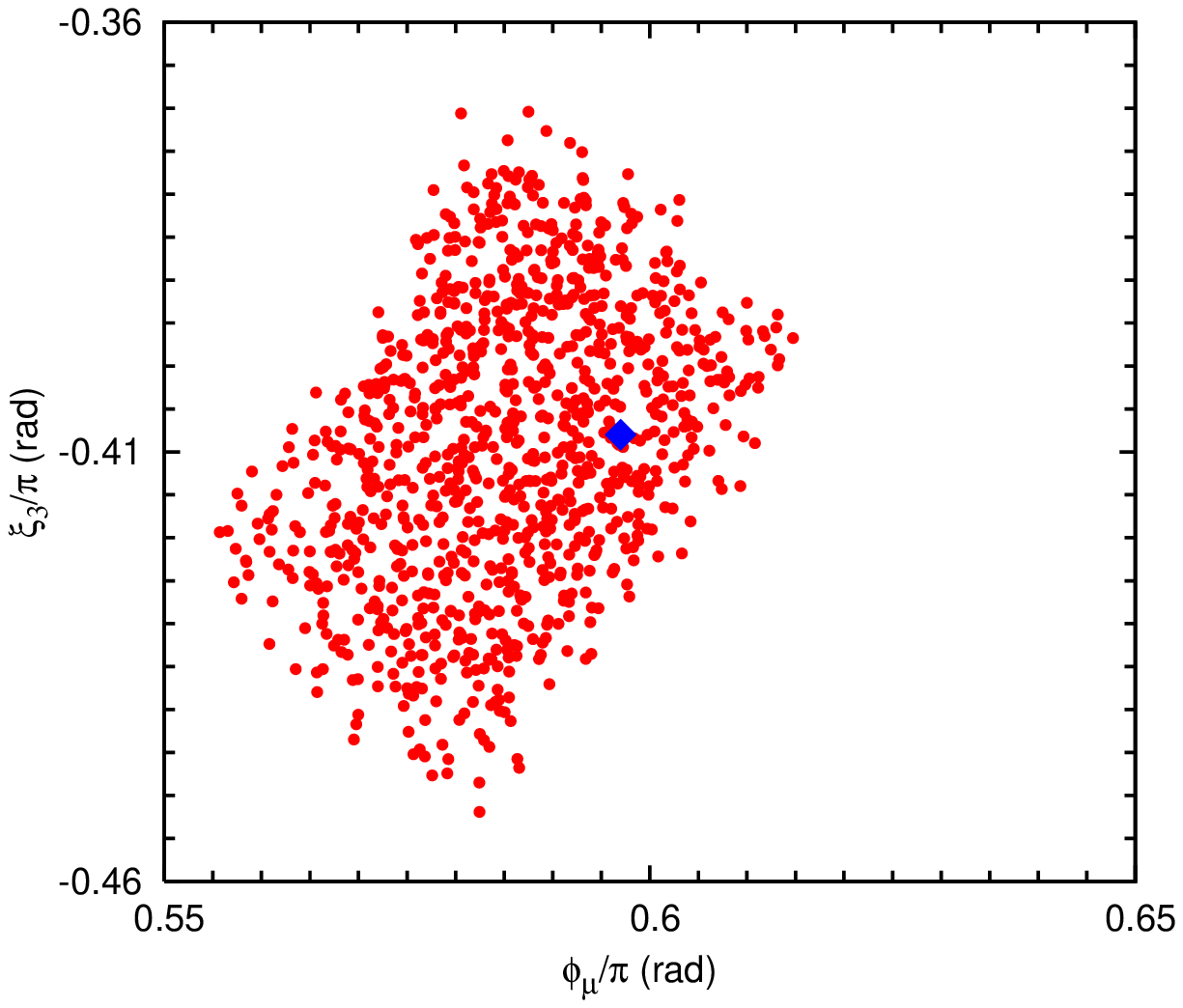}
\end{center}
\caption[]{Scatter plots  based on a random sample with 
 $m_0=2000\GeV$, $M_{1/2}=1720\GeV$, $A_0=100 \GeV$, $\tanb=45$
 and the phases in the region $-0.87<\xi_1/\pi<-0.77$,  $0.34<\xi_2/\pi<0.44$,
 $-0.46<\xi_3/\pi<-0.36$ and $0.55<\phi_\mu/\pi<0.65$. The dots (in red) are 
points of the random sample that satisfy all the EDM and cosmological constraints.
The (blue) diamond marks the values of the phases used  in the right panel of 
 Fig.~\ref{fig11}. }
\label{fig12}
\end{figure}
%%%%%%%%%%%%%%%

In order to show how sensitive the selected regions are to variations of the phases, which were chosen
to suppress the EDMs via the cancellation mechanism,
we pick a particular point located in the allowed funnel region on the left panel of Fig.~\ref{fig11}.
Then by varying the phases $\xi_{1,2,3}$ and $\phi_\mu$ around the values 
displayed in this figure, we produce a random sample consisting  of 100,000 points.
The scatter plots displayed in Fig.~\ref{fig12}, are based on this random sample with 
$m_0=2000\GeV$, $M_{1/2}=1720\GeV$, $A_0=100 \GeV$, $\tanb=45$
and  phases in the region $-0.87<\xi_1/\pi<-0.77$,  $0.34<\xi_2/\pi<0.44$,
 $-0.46<\xi_3/\pi<-0.36$ and $0.55<\phi_\mu/\pi<0.65$. The scattered dots ( in red ) represent the 
 random points  that satisfy all the EDM bounds and the WMAP3 bound on the 
 neutralino relic density. The (blue) diamond in the center of each panel, marks the point 
 corresponding to the values of phases on the right panel of 
 Fig.~\ref{fig11}. One can see that the allowed variations on  
 $\phi_\mu$ and $\xi_3$ are of the order of $\sim 0.05\pi$ rad. The same applies to the phase $\xi_2$ as well.
 On the other hand, the range of the values of the phase  $\xi_1$,  
 that are compatible with the EDM and cosmological constraints,  
 appears to be much broader.  Therefore, in the large $\tanb$ regime, the amount of tuning required 
 to locate  extended cosmological funnels that are also compatible 
 with the EDM bounds, is of the order of  $0.01\pi$ rad for $\xi_{2,3}$ and $\phi_\mu$
 and  $0.1\pi$ rad for $\xi_1$. 
In the displayed results the phases of the
trilinear couplings at the GUT scale are taken zero. However we have checked numerically 
that variations on them of the order $0.1\pi$ rad do not destabilize the cancellation
mechanism for the EDMs and therefore the fine tuning of these phases is less restrictive.

%%%%%%%%%%%%%%%%%
\section{Conclusions}
We have considered  supersymmetric 
models in the presence of CP-violating sources residing in the Higgsino-mixing mass term and 
SUSY breaking soft parameters. 
In the simple, mSUGRA based, versions of such models, that have been extensively studied 
in the past,  there are two  independent phases at the unification scale usually chosen to be the phase 
of the Higgsino mixing parameter $\mu$, $\phi_\mu$, and the phase of the 
common  trilinear coupling $A_0$, $\phi_A$. 
The application of the EDM and cosmological bounds constrains
the phase of $\mu$ to be $\phi_\mu/\pi \lesssim  0.01$ rad, whose smallness  poses severe 
obstacles in certain Baryogenesis mechanisms, while $\phi_A$ remains practically unconstrained and can be at least ten times larger. 
 
In our study we have used the revised top mass, whose central value is $m_t=171.4 \GeV$, and scan the parameter space allowing for non-universal boundary conditions for the phases at the GUT scale.  Such models are described by thirteen CP-violating phases,  
residing in the gaugino mass terms, the trilinear scalar couplings and the $\mu$ 
parameter, one of which can be rotated away. An additional phase which misaligns the VEVs of the Higgs doublets is determined from the minimization conditions and affects the analysis. 
We follow a top-bottom approach according to which the low energy values of all parameters involved, including their magnitudes and phases,  
are determined from their values at the  unification scale after  
running of the renormalization group equations. In such an analysis 
we showed that the two-loop running of the CP-violating phases induces 
important corrections to the electric dipole moments of the known species, which are absent in the one-loop running analysis.  This results to further constraints of the allowed parameter space. 
Important corrections to EDMs are also induced by the one-loop 
misalignment angle of the Higgs vacuum expectation values which are augmented 
for large $\tan \beta $ and in the presence of a non-vanishing gluino phase.   

The role of the gaugino phases is particularly important allowing for 
suppression of electric dipole moments if they are fine tuned at the unification scale.  
By tuning appropriately the  gaugino phases 
$\xi_{1,2,3}$ and the phase $\phi_\mu$ of $\mu$  
at the unification scale, there can be found   
regions in the parameter space, along the focus point and neutralino pair annihilation through a Higgs resonance, which are permitted by dark matter WMAP3 cosmological constraints,
and are also compatible with the electric dipole moment limits and all accelerator data. 

This can be accomplished for large  values  for the phases 
of the gaugino masses  and the $\mu$ parameter, of the order 
of the ${\cal{O}}(1)\;\pi$ rad, which are however fine-tuned.   
The phases $\xi_{2,3}$ and $\phi_\mu$ need to be adjusted  
at the $0.01\pi$ rad level, while the adjustment of $\xi_1$ is by an order of magnitude less restrictive. 
The phases of the trilinear scalar couplings although they play a key role for EDMs are not fine tuned when the magnitude of the trilinear coupling is small. 
The role of the gluino phase is important in this analysis. It can be used, by implementing the cancellation mechanism, to make neutron's electric dipole moment small, and in addition it  allows for a large non-vanishing phase for $\mu$. 
Large values of $\mu$ and gaugino phases are needed in certain 
electroweak Baryogenesis mechanisms . Higgsino and gaugino driven Baryogenesis 
requires the magnitude of $\mu$ to be comparable to the gaugino masses which can be fulfilled in the major portion of the parameter space in the constrained models considered in this work. 

Since the suppression of the
EDM is feasible for  small and intermediate values of the SUSY breaking scale, 
there are regions in the parameter space which are within the reach of the
forthcoming LHC experiments.

%%%%%%%%%%%%%%%%%%%%%%%%%%%%%%%%
\section*{Acknowledgments}
\noindent 
The work of M.A.  and A.B.L. was supported by funds made available by the 
European Social Fund (75\%)
and National (Greek) Resources (25\%) -- EPEAEK~B -- PYTHAGORAS. They  acknowledge also
support from the Special Research Account of the University of Athens.
The work of V.C.S.  was supported by Marie Curie Excellence grant   MEXT-CT-2004-014297 and 
Marie Curie International Reintegration grant``SUSYDM-PHEN",  MIRG-CT-2007-203189. 
In addition, A.B.L. and V.C.S.   acknowledge support by the Research Training Network 
``HEPTOOLS",  MRTN-CT-2006-035505. 

\vspace*{1cm}
%%%%%%%%%%%%%%%%%%%%%%%%%%%%%%%%%%%%%%%%%%%%%%%%%%%%%%%%%%%%%%%%%%%%%%%%%%%%%%%%%%%%%%%
%\newpage
%\begin{appendix}
\setcounter{section}{0}
\section{Appendix}
\setcounter{equation}{0}
\renewcommand{\theequation}{A.\arabic{equation}}

In this section we present the basic expressions and conventions used in this work so that a direct comparison with other works and different notations is made possible. 
The supersymmetric part of the Lagrangian is specified by a superpotential given by
%%%%%%%%%
\bea
{\cal W} \;=\; h_t \;Q^T \; \epsilon\; H_2\;U^c \;+\;
               h_b \;Q^T \; \epsilon\; H_1\;D^c \;+\;
               h_\tau \;L^T \; \epsilon\; H_2\;E^c \;+\;
               \mu \; H_1^T \; \epsilon\; H_2 \; , 
\label{sup}
\eea
%%%%%%%%%%
where the elements of the antisymmetric $2 \times 2$ matrix $\epsilon$ are
given by $\; \epsilon_{12}\;=\;-\epsilon_{21}\;=\;1\;$. In the
superpotential above we have only shown the dominant Yukawa terms of the
third generation and we do not allow for flavour mixings. According to this the Yukawa  
couplings of the left, right handed top and bottom multiplets are 
$$
{\cal W} \;=\; h_t \;H_2^0\; U\; U^c - h_b \;H_1^0\; D\; D^c \quad .
$$
%%%%%%%

The scalar soft part of the Lagrangian is given by
\bea
{\cal L}_{scalar} \;=\; &-&\;{\sum_i}\;m_i^2 \; |{\phi_i}|^2\; \nonumber \\ 
&-&\;(A_t\;h_t \;Q^T \; \epsilon\; H_2\;U^c \;+\;
    A_b\;h_b \;Q^T \; \epsilon\; H_1\;D^c \;+\;
    A_\tau\;h_\tau \;L^T \; \epsilon\; H_2\;E^c \;+\;h.c. ) \nonumber \\
    &-&\;(\; m_3^2\;H_1^T \; \epsilon\; H_2\;+\;h.c. )  \; ,
\eea
where the index $\;i\;$ in the sum in the equation above runs over all
scalar fields and all fields appearing denote scalar parts of the
supermultiplets involved.

The gaugino fields soft mass terms are given by
\bea
{\cal L}_{gaugino}\;=\; -{\frac{1}{2}}\;(\;M_1\;{\tilde B}\;{\tilde B}\;+\;
M_2\;{\tilde W}^{(i)}\;{\tilde W}^{(i)}\;+\;M_3\;{\tilde G}\;{\tilde G}\;
+\;h.c.) 
\eea
In this equation $\;{\tilde B},\;{\tilde W}^{(i)},\;{\tilde G}\;$ are the
gauge fermions corresponding to the $\;U(1),\;SU(2),\;$ and $\;SU(3)\;$ gauge
groups.
Our notation is that of Ellis and Zwirner \cite{Ellis:1989jf} with the signs 
of the gaugino masses 
and that of the bottom and tau Yukawa couplings reverted. The one-loop RGEs found 
in that reference coincide with ours if the signs of the gaugino masses are flipped. 
RGEs of course are insensitive to the Yukawa sign convention.  
%%%%%%%%%

The two-loop renormalization group equations (RGEs) in the general case, including 
supersymmetric CP violations, can be read from \cite{Martin:1993zk} and \cite{Yamada:1994id}. 
In \cite{Martin:1993zk} the soft SUSY breaking part of the Lagrangian is written as 
%%%%%%%
\beq
{\cal L}\;=\; - \frac{1}{6} \; h^{ijk} \;\phi_i \phi_j \phi_k - 
\frac{1}{2} b^{ij} \;\phi_i \phi_j - \frac{1}{2} \; {(m^2)}^j_i \phi_i {\phi^*}_j 
- \frac{1}{2} M_{(a)} \lambda_a \lambda_a + h.c.
\label{mart}
\eeq
%%%%%%
Keeping only flavour diagonal mass terms for the scalars 
$\phi_i$, as we have assumed throughout,  the mass terms become  
$$
{\cal L}\;=\;-Re{(m^2)}_i^i \; {| \phi_i|}^2 \,.
\label{massscalars}
$$
From Eq.~(\ref{massscalars}) we observe that the imaginary parts of ${(m^2)}_i^i $ do not 
appear in the Lagrangian. Besides it is easy to see that the RGEs of the quantities 
appearing in the Lagrangian do not depend on the imaginary parts of ${(m^2)}_i^i $. 
Therefore they do not affect the physical quantities and can be taken anything. 

For the two-loop RGEs, those of the trilinear couplings $A_{t,b,\tau}$ are found if one 
replaces 
$h_{ijk}$ and the Yukawas couplings $Y_{ijk}$ in the RGEs found in  \cite{Martin:1993zk} by 
$h \equiv -A \cdot Y$, where we have suppressed the flavour indices, and identifying the 
gaugino masses used in that reference with ours.  
The one-loop parts of the RGEs we get with such a replacement coincide with our one-loop 
RGEs for the $A_{t,b,\tau}$ which are identical, modulo the sign difference in the gaugino 
masses as mentioned before, with  reference \cite{Ellis:1989jf}. Note that the same would 
not have happened have we used $h \equiv -A \cdot Y$. To complete the correspondence and 
having as our guideline the one-loop results for the RGEs, $\mu$ of \cite{Martin:1993zk} 
should be replaced by our $\mu$ and the coupling $B$ of the two Higgs scalars should be 
replaced by our $-m_3^2$. With this correspondence the two-loop RGEs are retrieved 
unambiguously from \cite{Martin:1993zk} and are adapted to our notation. 

Concerning the neutralino and chargino mass matrices, 
In the $\tilde{B}$, $\tilde{W}^{(3)}$, 
$i \tilde{H}_{1}^0$, $i \tilde{H}_{2}^0$, basis 
the neutralino mass matrix is
%%%%%%%%%%%%%%%%%%%
\begin{equation}
{\cal M}_N \ =\ \left(\begin{array}{cccc} M_1 & 0 &
{g^\prime \vev{H_1^*}}   &
-{g^\prime  \vev{H_2^*}}
\\[1mm] 0 & M_2 & -{g \vev{H_1^*}} &
{g \vev{H_2^*}}
\\[1mm] {g^\prime  \vev{H_1^*}}&
-{g \vev{H_1^*}} & 0 & -\mu \\[1mm]
-{g^\prime  \vev{H_2^*}}
&{g \vev{H_2^*}} & -\mu & 0
\end{array} \right)\ .\label{neutra}
\end{equation}
%%%%%%%%%%%%%%
The field dependent neutralino mass matrix 
entering into the effective potential has a similar form with the Higgs VEVs replaced by the 
neutral components of the corresponding Higgs doublets. Note that the gaugino masses 
$M_{1,2}$ the parameter$\mu$ as well as the Higgss VEVs are complex in general. Regarding 
the Higgses one can verify that the mass eigenstates depend only on their relative phase. 
One can 
diagonalize the symmetric neutralino mass matrix as 
%The mass eigenstates (${\tilde \chi}_{1,2,3,4}^0$)
%%%%%%%%%%%%
\begin{equation}
{\cal O}^T {\cal M}_N {\cal O}\ ={\rm diag} \left (
m_{{\tilde \chi}^0_1},m_{{\tilde \chi}^0_2}, m_{{\tilde \chi}^0_3},
m_{{\tilde \chi}^0_4} \right ) \,,
\end{equation}
%%%%%%%%%%%%
where ${\cal O}$ is complex in general. In our approach this matrix is chosen is such a
 way that 
the eigenvalues $m_{{\tilde \chi}^0_i}$ are real and positive. 
%%%%%%%%%%%%

\noindent
The chargino mass matrix can be obtained from the Lagrangian terms
%%%%%%%%%%%%
\begin{equation}
{\cal L}^{mass}_{charginos} \ =\ 
- \left ( \tilde{W}^- , {i \tilde{H}_{1}^-} \right ) {\cal M}_c 
\left (\begin{array}{c} \tilde{W}^+ \\ {i \tilde{H}_{2}^+} \end{array}
\right ) \, + \, (h.c) \,,
\end{equation}
%%%%%%%%
where we have defined $\tilde{W}^\pm \equiv \frac{ \tilde{W}^{(1)}
\mp i \tilde{W}^{(2)} }{\sqrt{2}}$. The mass matrix is 
%%%%%%%%%
\begin{equation}
{\cal M}_C \ =\ \left (\begin{array}{cc} M_2 &
 -\;g\; \vev{H_2^*} \\[1mm]
-\;g\; \vev{H_1^*}& \mu \end{array}
\right )\, ,
\label{chmat}
\end{equation}
%%%%%%%%%%%%
which can be diagonalized as 
%%%%%%%%
\begin{equation}
U {\cal M}_c V^\dagger \ =\ \left (\begin{array}{cc} m_{\tilde{\chi}_1}
& 0 \\[1mm]
0 & m_{\tilde{\chi}_2} \end{array} \right ) \; .
\end{equation}
%%%%%%%%%%%%%%%%%%%
Thus,
%%%%%%%%%%%%
\begin{equation}
{\cal L}^{mass}_{charginos} \ =\ -m_{{\tilde{\chi}}_1} 
\bar{{\tilde{\chi}}}_1 \tilde{\chi}_1 -  m_{{\tilde{\chi}}_2}
\bar{{\tilde{\chi}}}_2 {\tilde{\chi}}_2 \; .
\end{equation}
%%%%%%%%%%%%%%
The Dirac chargino states $\tilde{\chi}_{1,2}$ in this equation are defined by
%%%%%%%%%%%%%%%
\begin{equation}
\tilde{\chi}_1 \equiv \left (\begin{array}{c} \lambda_1^+ \\
\bar{\lambda}_1^- \end{array} \right ) \,\, , \,\,
\tilde{\chi}_2 \equiv \left (\begin{array}{c} \lambda_2^+ \\
\bar{\lambda}_2^- \end{array} \right ) \,,
\end{equation}
%%%%%%%%%%%%%%%%%
with the two component Weyl spinors $\lambda^\pm_{1,2}$ related
to $\tilde{W}^\pm$, ${i \tilde{H}_{1}^-}$, ${i \tilde{H}_{2}^+}$ by
%%%%%%%%%%%%%%%
\begin{equation}
V \left ( \begin{array}{c} \tilde{W}^+ \\ {i \tilde{H}_{2}^+}
\end{array} \right ) \equiv \left (\begin{array}{c}
\lambda_1^+ \\ \lambda_2^+ \end{array} \right ) \,\, , \,\,
\left ( \tilde{W}^- \, , \, {i \tilde{H}_{1}^-} \right ) U^\dagger
 \equiv \left ( \lambda_1^- \, , \, \lambda_2^- \right ) \,.
\end{equation}
%%%%%%%%%%%%%%%%

The gauge interactions of charginos and neutralinos can
be read from the  Lagrangian\footnote{
In our notation ${e} \equiv $electron's charge.} 
%%%%%%%%%%%%%%%
\begin{equation}
{\cal L} \ =\ {g} \left ( W^+_\mu J_-^\mu + W_\mu^- J_+^\mu 
\right ) + {e} A_\mu J_{em}^\mu +\frac{{e}}{{s} {c}}
Z_\mu J_Z^\mu \,.
\end{equation}
%%%%%%%%%%%%%%%%%
In the equation above $\;s\;=\; \sin \theta_W,\;c\;=\; \cos \theta_W\;$.
Also,
%%%%%%%%%%%%%
\begin{equation}
\left (\begin{array}{c} Z_\mu \\[1mm] A_\mu \end{array} \right )
\ =\ \left ( \begin{array}{cc} {c} & {s} \\[1mm]
-{s} & {c} \end{array} \right ) \, 
\left ( \begin{array}{c} W_\mu^{(3)} \\[1mm] B_\mu \end{array}
\right ) \,.
\end{equation}
%%%%%%%%%%%%%%%%%
The currents $J_+^\mu$, $J_{em}^\mu$ and $J_Z^\mu$ are given by
%%%%%%%%%%%%%%%%%
\begin{equation}
J_+^\mu \equiv {\bar {{\tilde \chi}}^0_a}  \gamma^\mu \left [
{\cal P}_L {\cal P}^L_{a i}+ 
{\cal P}_R {\cal P}^R_{a i}  \right ]
\tilde{\chi}_i \;\quad  a=1...4,\;\; i=1,2 \,,
\end{equation}
%%%%%%%%%%%%%%
where
${\cal P}_{L,R} = \frac{1 \mp \gamma_5 }{2}$ and
%%%%%%%%%%%%%%%
\begin{eqnarray}
{\cal P}^L_{a i}\equiv +\frac{1}{\sqrt{2}} {\cal O}_{4 a}^{*} V^{*}_{i 2}
- {\cal O}_{2 a}^{*} V^{*}_{i 1}\quad , \quad
{\cal P}^R_{a i}\equiv -\frac{1}{\sqrt{2}} {\cal O}_{3 a} U^{*}_{i 2}
- {\cal O}_{2 a} U^{*}_{i 1} \,.
\end{eqnarray}
%%%%%%%%%%%%%% 
The electromagnetic current $J_{em}^\mu$ is
%%%%%%%%%%%%
\begin{equation}
J_{em}^\mu \ =\ \bar{\tilde\chi}_1 \gamma^\mu \tilde{\chi}_1 +
\bar{\tilde\chi}_2 \gamma^\mu \tilde{\chi}_2 \,.
\end{equation}
%%%%%%%%%%%%%%%%%%%
Finally, the neutral current $J_Z^\mu$ can be read from
%%%%%%%%%%%%%%%
\begin{equation}
J^\mu_Z \equiv {\bar{{\tilde \chi}}}_i  \gamma^\mu \left [\;
{\cal P}_L {\cal A}^L_{i j} + {\cal P}_R {\cal A}^R_{i j} \; \right ]
\tilde{\chi}_j + \frac{1}{2} \; {\bar {{\tilde \chi}}}^0_a \gamma^\mu \left [\;
{\cal P}_L {\cal B}^L_{a b} + {\cal P}_R {\cal B}^R_{a b} \;\right ]
{{\tilde \chi}^0_b} \,,
\end{equation}
%%%%%%%%%%%%%%%%%%%%%
with
%%%%%%%%%%%%%%
\begin{eqnarray}
{\cal A}^L_{i j} &=& {{c}^2} \delta_{i j} -\frac{1}{2} V_{i 2} V^{*}_{j 2}
\,,\nonumber \\
{\cal A}^R_{i j} &=& {{c}^2} \delta_{i j} -\frac{1}{2} U_{i 2} U^{*}_{j 2}
\,,\nonumber \\
{\cal B}^L_{a b} &=& \frac{1}{2} \left ( {\cal O}_{3 a}^{*} {\cal O}_{3 b} -
{\cal O}_{4 a}^{*} {\cal O}_{4 b} \right )
\,,\nonumber \\
{\cal B}^R_{a b} &=& - {\cal B}^{L,*}_{a b} \,.
\end{eqnarray}
%%%%%%%%%%%%%%%%%%

The chargino and neutralino couplings to sfermions 
are given by the following Lagrangian terms
%%%%%%%%%%%%%
\begin{equation}
{\cal L} = i \; {\bar{\tilde {\chi}}}_{i}^{c} \;
 ({\cal P}_L \, a^{ {f^{\prime}}   {\tilde f} }_{ij} +
{\cal P}_R \, b^{ {f^{\prime}}   {\tilde f} }_{ij}) \,  {f^\prime } \,
 {\tilde f}_{j}^{*}
\, + \,
i \; {\bar{\tilde {\chi}}}_{i} \;
 ({\cal P}_L \, a^{f {\tilde f}^{\prime} }_{ij} +
{\cal P}_R \, b^{f {\tilde f}^{\prime} }_{ij}) \,  {f} \,
 {\tilde f}_{j}^{\prime *} \, + \, (h.c) \,.
\end{equation}
In this, ${\chi}_{i} \; (i=1,2)$ are the positively charged charginos  and 
${\chi}_{i}^{c}$ the corresponding charge conjugate states having 
opposite charge. $f\, , \,{f}^{\prime}$ denote ``up" and ``down"
fermions, quarks or leptons, while ${\tilde f}_i\, , \,{\tilde f}_i^{\prime}$
are the corresponding sfermion mass eigenstates.
The left and right-handed couplings appearing above are given by
%%%%%%%%%%%%%%%%%%%%%%
\begin{eqnarray}
a^{{f^{\prime}}{\tilde f} }_{ij} \ori &= \ori
g V_{i1}^{*} \, K^{\tilde f}_{j1} - h_{f} V_{i2}^{*}  K^{\tilde f}_{j2} \ori ,
\ori  &b^{ {f^{\prime}} {\tilde f} }_{ij} \ori =
 \ori -h_{f ^\prime} \,  U_{i2}^{*} K^{\tilde f}_{j1}  \,,  \nonumber  \\
a^{f {\tilde f}^{\prime} }_{ij} \ori &= \ori
g U_{i1} \, K^{{\tilde f}^{\prime}}_{j1} + h_{f^\prime} \, U_{i2}
  K^{{\tilde f}^{\prime}}_{j2} \ori , \ori
&b^{f {\tilde f}^{\prime} }_{ij} \ori = \ori
h_{f } \,  V_{i2} K^{{\tilde f}^{\prime}}_{j1} \,.  \nonumber
\end{eqnarray}
%%%%%%%%%%%%%%%%%%%%%%
In the equation above $h_{f } \, , \, h_{f ^\prime} $ are the Yukawa
couplings of the up and down fermions respectively. The matrices
$K^{\tilde{f},{\tilde f}^\prime} $ which diagonalize the sfermion mass matrices become the
unit matrices in the absence of left-right sfermion mixings.
For the electron and muon family the lepton masses are taken to be
vanishing in the case that mixings do not occur. In addition the 
right-handed couplings, are zero.  \newline 
The corresponding neutralino couplings are given by
%%%%%%%%%%%%%%%%%%%%%%%%%%%%%%%%%%%%
\begin{equation}
{\cal L} = i \;{\bar{\tilde {\chi}}}_{a}^{0}
 ({\cal P}_L \, a^{ {f}   {\tilde f} }_{aj} +
{\cal P}_R \, b^{ {f}   {\tilde f} }_{aj}) \,  {f} \,
 {\tilde f}_{j}^{*}
\, + \,
i \; {\bar{\tilde {\chi}}}_{a}^{0} \;
 ({\cal P}_L \, a^{{f^{\prime}} {\tilde f}^{\prime} }_{aj} +
{\cal P}_R \, b^{{f^{\prime}} {\tilde f}^{\prime} }_{aj}) \, {f^{\prime}} \,
 {\tilde f}_{j}^{\prime *} \, + \, (h.c) \,.
\end{equation}
%%%%%%%%%%%%%%%%%%%%%%%
The left and right-handed couplings for the up fermions, sfermions
are given by
%%%%%%%%%%%%%%%%%%%%%%
\begin{eqnarray}
a^{{f}{\tilde f}}_{aj} \ori &= &\ori
{\sqrt{2}} \, ( g {T^{3}_f}{O_{2a}}+{g^{\prime}}{\frac{Y_f}{2}} \,{O_{1a}})
\, {K^{f}_{j1}} \ori + \ori h_{f} \, {O_{4a}}\,  {K^{f}_{j2}}  \ori , \ori
\nonumber \\
b^{{f}{\tilde f} }_{aj} \ori & = & \ori
{\sqrt{2}} \, (-{g^{\prime}}{\frac{Y_{f^c}}{2}} \,{O_{1a}^{*}}) \, {K^{f}_{j2}}
\ori - \ori h_{f} \, {O_{4a}^{*}}\,  {K^{f}_{j1}} \,,  \nonumber  
\end{eqnarray}
%%%%%%%%%%%%%%%%%%%%%%
while those for the down fermions and sfermions are given by
%%%%%%%%%%%%%%%%%%%%%%%%%%%%%
\begin{eqnarray}
a^{{\fp}{\tilde \fp}}_{aj} \ori &= &\ori
{\sqrt{2}} \,
( g {T^{3}_{\fp}}{O_{2a}}+{g^{\prime}}{\frac{Y_{\fp}}{2}} \,{O_{1a}})
\, {K^{\fp}_{j1}} \ori - \ori h_{\fp} \, {O_{3a}}\,  {K^{\fp}_{j2}}  \ori , \ori
\nonumber \\
b^{{\fp}{\tilde \fp} }_{aj} \ori & = & \ori
{\sqrt{2}} \, (-{g^{\prime}} \frac{Y_{ {\fp}^c }}{2} \,{O_{1a}^{*}})
 \, {K^{\fp}_{j2}}
\ori + \ori h_{\fp} \, {O_{3a}^{*}}\,  {K^{\fp}_{j1}}\,.   \nonumber  
\end{eqnarray}
%%
%\end{appendix}
%%%%%%%
%%%%
\clearpage
%%
%%%%%%%%%%%%%%%%%%%%%%%%%%%%%%%%%%%%%%%%%%%%%%%%%%%%%%%%%%%%%%%%%%%%%
%\newpage
%%%%

%%%%%%%%%%
\end{document}